\documentclass{aa}  

\usepackage{graphicx}
\usepackage{xcolor}
\usepackage{txfonts}
\usepackage{xspace}
\usepackage{amsmath,mathrsfs,amssymb,amsbsy,bm}

\newcommand{\AMD}{\ensuremath{\mathrm{AMD}}\xspace}
\newcommand{\AMDe}{\ensuremath{\mathrm{AMD}_{\mathrm e}}\xspace}
\newcommand{\AMDi}{\ensuremath{\mathrm{AMD}_{\mathrm i}}\xspace}
\newcommand{\AMDH}{\ensuremath{\mathrm{AMD}_{\mathrm H}}\xspace}
\newcommand{\Ms}{\ensuremath{\mathrm{M}_\odot}}
\newcommand{\Me}{\ensuremath{\mathrm{M}_\oplus}}
\newcommand{\std}{\ensuremath{\mathrm{std}}}
\newcommand{\Var}{\ensuremath{\mathrm{Var}}}

\usepackage{comment}
\usepackage[colorlinks=true,linkcolor=blue,citecolor=blue,urlcolor=blue]{hyperref}
\usepackage{multirow}
\usepackage{tablefootnote}
\usepackage{natbib}
\usepackage{subcaption}         %
\usepackage{lscape}             %
\usepackage{placeins}           %

\begin{document}

   \title{Eccentricities and inclinations are decoupled in super-Earth systems}

   \author{S. Kuhn \inst{1,2} \corrauth{sacha.kuhn@oca.eu}
        \and A. C. Petit \inst{1} \email{antoine.petit@oca.eu} \and  F. Mogavero\inst{3} \email{mogavero@math.unipd.it} \and A. Crida\inst{1} \email{aurelien.crida@oca.eu}
        }

   \institute{Université Côte d’Azur, Observatoire de la Côte d’Azur, CNRS, Laboratoire Lagrange, France
            \and MAUCA -- Master track in Astrophysics, Université Côte d'Azur \& Observatoire de la Côte d'Azur, Parc Valrose, 06100 Nice, France
            \and Università degli Studi di Padova, Dipartimento di Matematica "Tullio Levi-Civita", Via Trieste 63, 35121 Padova, Italy\\ }

   \date{Received September 30, 20XX}

\abstract
{
Recent work on the longterm evolution of the inner Solar System has shown that the inclination contribution to the Angular Momentum Deficit (AMD) is almost constant even though inclinations and eccentricities are highly coupled in analytical expansions of the perturbing function.
}
{
We wish to characterize the conservation of the eccentricity AMD and inclination AMD, respectively describing the average eccentricity and mutual inclination of planetary orbits, beyond the case of the Solar System. We aim at showing that this conservation is a general feature of the stable planetary systems dynamics, in particular, in the context of low mass multiplanetary extrasolar systems.
}
{We perform a series of numerical simulations using both N-body and secular evolution codes to study planet dynamics over billions of orbits.
Taking as a template four equal mass Super-Earth systems with non uniform orbital spacing, we consider sets of systems with different initial AMD and use the orbital elements evolution for each planet to assess the conservation of both inclination and eccentricity AMD components.
}
{For stable systems, the inclination AMD and eccentricity AMD relative variation is typically below 1\%, making them  quasi-integrals of motion. At low AMD, despite short Lyapunov timescales, the secular diffusion is confined and the dynamics almost quasi-periodic. As the AMD increases, the emergence of large scale secular chaos only weakly affects the conservation of the quasi-integrals, which only breaks when systems become fully unstable.}
{The eccentricity and inclination evolution (as measured by the variation of their AMD contributions) are largely decoupled for low mass multiplanetary systems.

Combined with the weak diffusion of the Laplace-Lagrange modes at low AMD, these findings paint a picture where the secular dynamics are not necessarily ergodic.
The uncovering of these quasi-integrals opens a new path to relate the observed planetary architecture to its formation.
}


\titlerunning{Quasi-integrals in secular dynamics}
\authorrunning{S. Kuhn et al.}

   \maketitle
\nolinenumbers
\section{Introduction}
\label{sec:Introduction}
Since the very first discovery of an exoplanet by \citet{Mayor_1995}, the astrophysics community has been actively looking to discover more and more exoplanets.
In this scope, many transit space-based missions have been completed \citep[see e.g.][for the CoRoT and Kepler missions]{Auvergne_2009,Borucki_2010} and many more are undergoing, as TESS \citep{Ricker_2014} and, in preparation, as PLATO \citep{Rauer_2025}. Radial velocity surveys have also allowed a large number of exoplanet discoveries, using for instance the spectrographs SOPHIE \citep{Bouchy_2013} and HARPS \citep{Mayor_2011, Bonfils_2013, Mignon_2024} or via the California Legacy Survey \citep{Rosenthal_2021}.
With no less than 6,451 confirmed exoplanets discovered to date\footnote{As of October 15, 2026, according to the NASA exoplanet archive available at \url{https://exoplanetarchive.ipac.caltech.edu/}.}, understanding their dynamics has been a focus.
There have indeed been major advances in our understanding of exoplanet population distributions and the structure of planetary systems.
It has become clear that the Solar System is highly atypical and does not follow the common architecture of planetary systems.
Indeed, many systems seem to be formed of closely-packed, similarly massive planets \citep{Weiss_2018} and have very low mutual inclinations ($\lesssim5^\circ$) \citep{Tremaine_2012} and eccentricities ($\lesssim0.1$) \citep{Winn_2015, Xie_2016, Johansen_2012}.

While much of the focus has been on the diversity and dynamical histories of exoplanetary systems, intriguing insights have also emerged from revisiting the dynamics of our own Solar System. 

The dynamical study of the Solar System has always been an active field of research and the first longterm simulations of the Solar System have shown its chaotic nature \citep[][]{Laskar_1989,Laskar_1990}.
As a result, it is impossible to accurately predict its evolution beyond about 60 million years (Myr). 
For longer timescales, it is still possible to make predictions, but through a statistical approach \citep{Laskar_2008}.
In \citet{Laskar_2009}, the probability that Mercury's orbit might cross the path of Venus was found to be about $1\%$ in the next 5 billion years.
This apparent discrepancy between the long stability timescale and the short Lyapunov timescale of only 5 Myr found by \citet{Laskar_1989} hints to more complex dynamical features.
 Recent studies on the inner Solar System dynamics \citep{Mogavero_2023} have shown that the ergodic assumption (under which the entire phase space of the system may be explored) is not verified for the secular chaos, with quantities being almost conserved over planetary systems lifetime.
In particular, an unexpected result was found by \citet{Mogavero_2023} regarding the longterm evolution of the Angular Momentum Deficit \citep[hereafter \AMD,][]{Laskar_1997,Laskar_2000}.
The \AMD is the difference between the norm of the angular momentum of a coplanar and circular system with the same semi-major axes and the norm of the angular momentum.
The AMD can be seen as, approximately, a weighted sum of the squared mutual inclinations and squared eccentricities of the planetary orbits.
\citet{Mogavero_2023} found that the part of the \AMD originating from the inclinations (\AMDi) is almost conserved in the inner Solar System\footnote{The inner Solar System \AMD is not conserved due to the forcing caused by the outer system inducing variations of \AMD originating from the eccentricities (\AMDe).}.
The conservation of \AMDi in the inner system is a possible explanation for the discrepancy between the Lyapunov timescale and the collision probability mentioned above.
In fact, a conservation is a very unexpected result since the variables are strongly influencing each-other and exchanges occur for very inclined systems, for instance, through the well known Kozai-Lidov oscillations \citep{Kozai_1962, Lidov_1962}.

The amount of AMD acts as a dynamical temperature and determines the possibilities for planets within a system to suffer from close encounters.
Having the \AMDi constant limits the pathways through which the system could evolve into an unstable configuration.
Thus, systems tend to be more stable when the \AMDi is perfectly conserved than when it is not.
Generalizing the conservation of \AMDi beyond the Solar System could be a new probe for understanding exoplanet formation conditions.
We know from the work of \citet{Stewart_1988}, \citet{Ohtsuki_2002}, and \citet{Stewart_2000} that, during instability phases, scattering of planetesimals in the disk tends to excite eccentricities faster than inclinations.
If the conservation of \AMDi found by \citet{Mogavero_2023} in the case of the Solar System is valid for exoplanetary systems, the current state of systems could be related to formation conditions within disks, and thus, for instance, provide insights into instability phases.
Finally, this conservation has an implication for understanding how exoplanetary systems evolve over long timescales but also constrains the occurrence rate of multi-planetary systems.
Indeed the detection capabilities of such systems are strongly impacted by the mutual inclination distribution (in particular in the case of the transit technique).

The results obtained in the 90s and 2000s, but also the recent work of \citet{Mogavero_2023}, were all made possible by a specific approach to the dynamics.
Outside Mean Motion Resonances (MMRs) active regions, the motion of planets around massive objects like stars is dominated by the slow
deformation of their orbits while keeping a constant semi-major axis.
It is thus possible to simplify the system's physics and drastically reduce the computation time of longterm evolution simulations.
These dynamics, called secular, are known since Laplace and Lagrange's times.

In the secular dynamics approach, one averages the equations of motion over an orbital period, and thus considers constant semi-major axes.
One can use either the Gauss numerical averaging method \citep{Gauss_1818} or the power series method \citep{Laskar_1989} that uses a development in power series of eccentricity and inclination.
In the former case, the averaging process is mathematically equivalent to replacing each planet with its instantaneous Keplerian orbit, treating the planet’s mass as continuously spread along the orbit in proportion to the time it spends at each point.
This results in a mass distribution that is inversely proportional to the orbital speed.
Secular dynamics then describes the slow gravitational interaction between these smeared-out Keplerian rings.
This method does not imply a development in eccentricity nor in inclination but is restricted to a simple averaging in time, which is only accurate to first order in mass \citep{Laskar_1989}.
On the other hand the power series method can be used with high order expansion in masses and is thus preferred in the Solar System secular dynamics, where a first order expansion simply does not mimic the true behavior of the system due to the large mass of the outer planets and the proximity to the 5:2 Jupiter-Saturn MMR.
The lowest degree expansion corresponds to the Laplace-Lagrange dynamics, for which the eccentricities and inclinations are fully decoupled and which yields quasi-periodic solutions to the orbital elements' evolutions. We will consider secular numerical integrations to first order in mass and without any expansion in eccentricities nor inclinations.
Integrations will go up to $10^{11}$ orbits, which is comparable to the systems' real ages and is larger than most other exoplanetary systems evolution studies (typical integrations of systems usually go up to about 100 Myr for formation studies).

However, the validity of the secular dynamics approach is not guaranteed for all systems.
It is only valid in the case where the system is sufficiently stable, i.e. outside of MMRs active regions.
Recent studies \citep{Izidoro_2017, Izidoro_2021} have shown that the majority of exoplanet systems are non resonant, ensuring the reliability of the secular dynamics approach.
Many studies have also been carried out to assess the longterm dynamics and stability of exosystems, in particular using machine learning with the work of \citet{Tamayo_2020}, or using analytic developments via the AMD stability \citep[see e.g.][]{Laskar_2017, Petit_2018}. 
Finally the secular dynamics approach on chaotic systems has also been extensively studied, for instance by \citet{Wu_2011}, \citet{Petrovich_2019} and \citet{Mogavero_2022} and has proved its reliability. 
Yet an analytical description of the chaotic diffusion of the secular dynamics remains a challenging task.
Indeed, since the early works of Chririkov \citep{Chirikov_1979}, most of the studies of chaotic diffusion processes focused on problems where the diffusion happens mostly along a one-dimensional manifold \citep{Petit_2020, Hussain_2020,Hadden_2024,Huang_2026}.
However, the secular dynamics involves a complex resonance network spanning a high dimensional space with no obvious preferred direction.

The longterm conservation effect of \AMDi was pointed out for the low eccentricities and inclinations of the inner Solar System planets.
However this conservation might be due to the architecture of the Solar System, which, as said before, does not follow the one of a typical planetary system.
The question then remains as to whether the conservation is relevant to other systems.
This paper constitutes a first study of the conservation of both \AMD components, i.e. the inclination part (\AMDi) and eccentricity part (\AMDe) contributing to the total \AMD, beyond the case of the Solar System. We will thus consider systems much different than the Solar System.
Since the Kepler mission has been discovering a large amount of very compact super-Earth multiplanetary systems \citep{Fabrycky_2014}, we will focus our analysis on such systems.
Another key point is that we will consider some systems where the innermost planet is close to the star, again typical of compact exoplanet systems.
The latter consideration will drastically affect the relativistic precession of planets. We also consider different amounts of mutual inclinations and eccentricities. Our goal is to investigate whether a conservation exists beyond the Solar System and, if so, to identify where the transition between conservation and non-conservation occurs.

This manuscript is organized as follows: in Sec. \ref{sec:phenomenology} we present a qualitative description of the different dynamical models for planetary dynamics, we then focus in Sect. \ref{sec:Secular_dynamics} on secular dynamics and present the Angular Momentum Deficit.
We introduce in Sect. \ref{sec:Numerical} the codes used for the simulations and the planetary systems considered.
We show our results on the existence of quasi-integrals in Sect. \ref{sec:Results} and finally Sect. \ref{sec:discussions} is dedicated to discussions over the results.
\section{Qualitative dynamical description}\label{sec:phenomenology}
\begin{figure}%
\begin{center}
\includegraphics[width=1.0\linewidth]{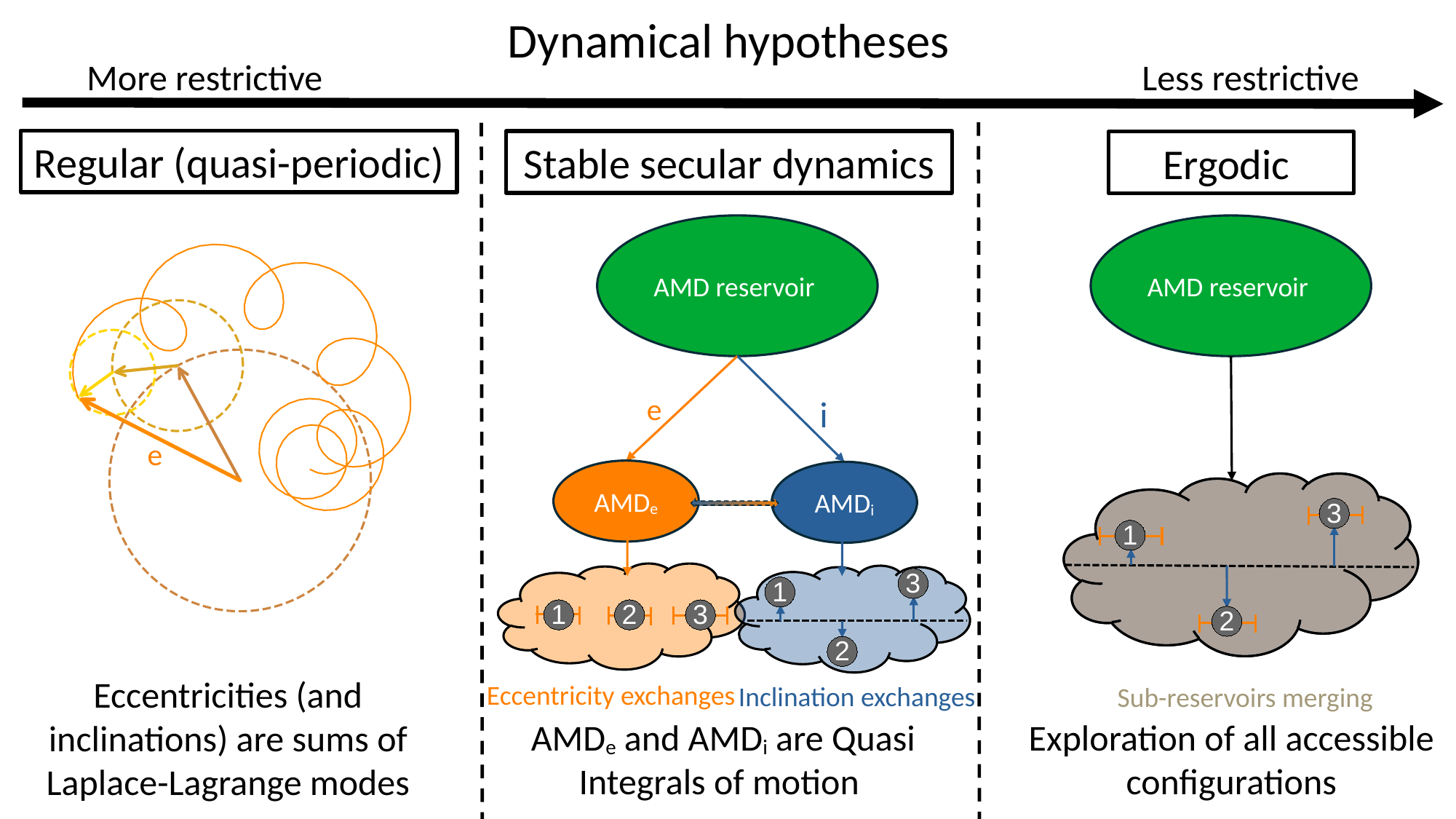}
\caption{Representation of the different dynamical hypotheses suitable to the dynamics of a planetary system (composed of 3 planets here).
The work presented in this paper focuses on the stable secular dynamics and the notion of quasi-integrals of motion.
In this framework, the AMD is represented split into the \AMDe and \AMDi components which hardly communicate.
These form two separate "sub-reservoirs" of excitation.
However, in the ergodic framework, the sub-reservoirs can be seen as merged, the dynamics can thus explore a wider range of configurations.}
\label{fig:sketch}
\end{center}
\end{figure}

We give here a high level description of the dynamics at play to fix the general concepts we will use in the paper.
The precise dynamical definitions can be found in Sect. \ref{sec:Secular_dynamics}.
The N-body problem being non-integrable, several approximations can be made to give a qualitative description of the dynamics.
We propose a simple representation of these models in Fig. \ref{fig:sketch} to fix ideas on the work presented here.
The Laplace-Lagrange dynamics is the most restrictive description of the dynamics.
Indeed, under this model, the evolution of eccentricities and inclinations are completely decoupled and evolve as superpositions of modes with different frequencies.
This model gives a good description of the dynamics in regions where MMRs are negligible and only at low levels of eccentricities and inclinations, or at higher levels but for short term integrations.
It simply gives a typical idea of the magnitude of possible variations of the eccentricity and inclination for each planet of the system.
This model is a significant approximation, it does not capture the complexity of the true dynamics.
At the other extreme, we could imagine that, on very long timescales, the dynamics follows an ergodic behavior, i.e. the system considered is not restricted by anything but the conservation of the energy and angular momentum integrals of motion and can explore the entire remaining phase space.
As mentioned earlier, the work of \citet{Mogavero_2023} shows that the ergodic assumption does not hold for the inner Solar System.
However, the probability density functions of the eccentricities and inclinations predicted by an ad-hoc ergodic ansatz in \citet{Mogavero_2017} have been shown to provide overall reasonable estimates for the longterm dynamics of the inner planets.
It is then obvious that there exists a tension between the true dynamics and this ergodic model.
This leads us to investigate a more intermediate description of the motion, the stable secular dynamics of planetary systems, where \AMDi and \AMDe are quasi-integrals.
These quasi-integrals effectively restrict the freedom of the system and splits the \AMD into two separate reservoirs.

\begin{figure}%
\begin{center}
\includegraphics[width=1.0\linewidth]{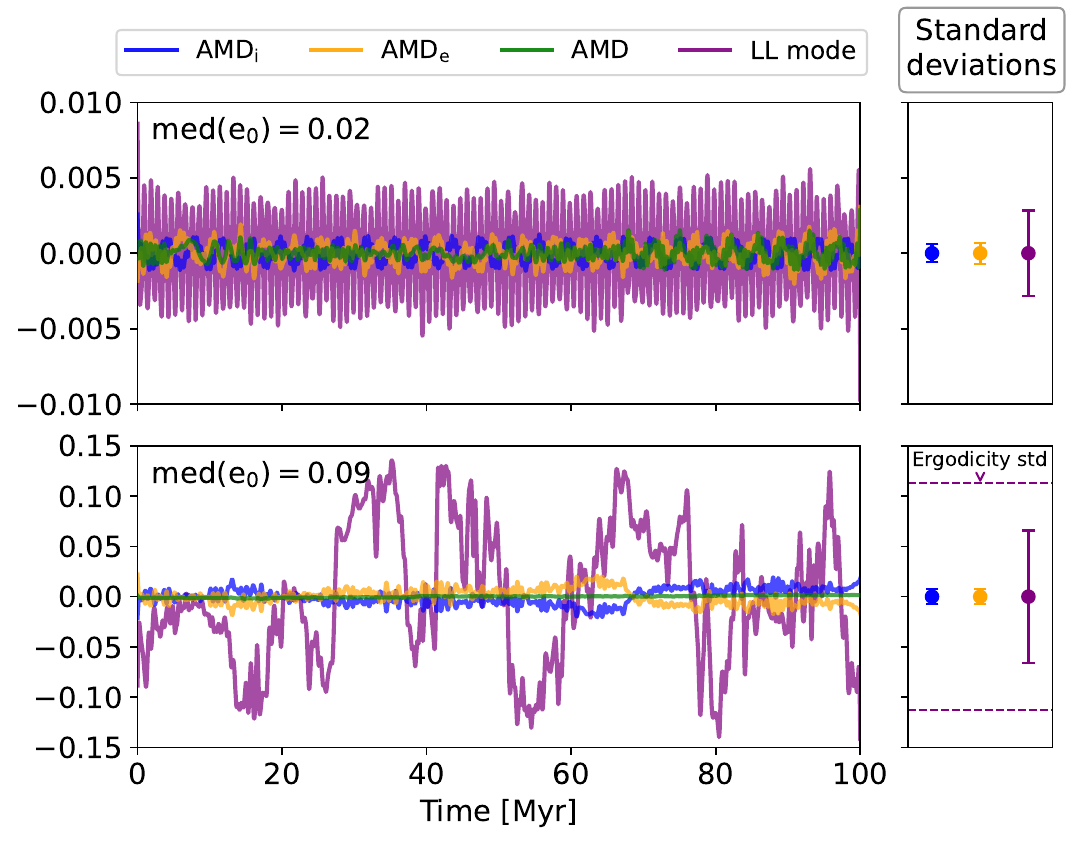}
\caption{Evolution of the \AMDi, \AMDe, \AMD and squared modulus of the eccentricity Laplace-Lagrange mode with largest variability, over 100 Myr for 2 different systems belonging to the setups \texttt{Peas}$_3$ (see Table. \ref{tab:Simulation_params} and Sect. \ref{sec:Numerical} for details) and \texttt{Peas}$_6$ (respectively top and bottom subplots).
Each quantity has the mean over 100 Myr subtracted and has been normalized by the temporal mean of the \AMD.
The time series are low-pass filtered using a cutoff frequency of 0.5 Myr via a 4$^{\text{th}}$ order Butterworth filter. Note that there is a change in the y scale between the two subplots. We have two different behaviors in the two different systems.}
\label{fig:AMD_mode_variability}
\end{center}
\end{figure}

We show in Fig. \ref{fig:AMD_mode_variability} the time series of the \AMDi, \AMDe, \AMD and an eccentricity proper mode over 100 Myr for two different 4-planet systems.
Both systems have a different amount of initial \AMD ($\AMD_0$) and thus have different dynamical behaviors. 
In the case of the low $\AMD_0$ system (top) we notice an underlying Laplace-Lagrange (LL) dynamics dominating the evolution.
We have clear small amplitude quasi-periodic variation of the \AMDi, \AMDe and consequently of \AMD.
Obviously the squared modulus of the proper mode does as well vary in a periodic-like manner, but with an amplitude about 4 times larger than \AMDi and \AMDe.
In the case of larger $\AMD_0$ (bottom), the variations of the proper mode have much bigger amplitudes than the \AMD components over the 100 Myr of evolution.
The \AMDi and \AMDe are still well conserved but vary much more than before. We can say that, in Fig. \ref{fig:AMD_mode_variability} (top), the dynamics is dominated by the LL model.
Indeed, in LL dynamics, the individual modes are by definition conserved and we notice in Fig. \ref{fig:AMD_mode_variability} (top) that the conservation of the individual mode is excellent.
The resonance network conserves \AMDi and hence negligible exchanges occur between the reservoirs.
However, Fig. \ref{fig:AMD_mode_variability} (bottom) is highly non-LL-driven and resonances tend to trigger changes in \AMDi and \AMDe.
In the next section, we propose a brief review of the Hamiltonian, AMD, Laplace-Lagrange and proper modes formalisms mentioned.

\section{Secular dynamics}\label{sec:Secular_dynamics}
We detail here the construction of the secular Hamiltonian considered for our study, as well as the notion of Angular Momentum Deficit. 
We remind the Laplace-Lagrange solutions in the secular dynamics framework and present the origin of the quasi-integral nature of the AMD components.

\subsection{Planetary Hamiltonian}
Let us consider $N+1$ bodies of masses $m_0$, $m_1$,$\dots$, $m_N$ and such that $m_0\gg m_k$, $k \in \{1,...N\}$ and with $\mathcal{G}$ the gravitational constant.

The Hamiltonian of this system, $\mathcal{H}$, is a sum of integrable 2-body problems, $\mathcal{H}_0$, with a perturbation, $\mathcal{H}_1$.
We use the classical complex Poincar\'e coordinates $x_k$ and $y_k$ \citep{Laskar_1995}, forming with $-\iota \bar{x}_k$ and $-\iota \bar{y}_k$ a basis of canonical coordinates.
They are defined as
\begin{equation}
\begin{aligned}
    x_k &= \sqrt{\Lambda_k} \sqrt{1 - \sqrt{1 - e_k^2}}\, \mathrm{E}^{\iota\varpi_k}, \simeq \sqrt{\Lambda_k/2}e_k\mathrm{E}^{\iota\varpi_k},\\
    y_k &= \sqrt{\Lambda_k} \sqrt{\sqrt{1 - e_k^2}(1 - \cos(i_k))}\, \mathrm{E}^{\iota\Omega_k} \simeq \sqrt{\Lambda_k/2}i_k\,\mathrm{E}^{\iota\Omega_k},
\end{aligned}
\label{eq:x_y_var}
\end{equation}
where\footnote{The choice of coordinate $\Lambda$ made here is called democratic heliocentric \citep{Hernandez_2017}.} $\Lambda_k = m_k \sqrt{\mathcal{G}m_0 a_k}$, $a_k$ are the planet semi-major axes, $e_k$ the eccentricities, $i_k$ the inclinations, $\varpi_k$ the longitudes of the perihelia and $\Omega_k$ the longitudes of the nodes. 
From now on, ${\iota = \sqrt{-1}}$ and $\mathrm{E}$ is the exponential operator.
The integrable part of the Hamiltonian is thus
\begin{equation}
    \mathcal{H}_0=-\sum_{k=1}^N\frac{\mathcal{G} m_0m_k^3}{2\Lambda_k^2},
    \label{eq:H_0}
\end{equation}
so that the Poincar\'e’s complex variables are integrals of motion\footnote{In other words, conserved quantities.} for the Kepler problem.
\subsection{Secular planetary Hamiltonian}
In the case where planets do not get too close to each other (i.e. do not suffer close encounters) and do not lie near MMRs, it is customary to work with an averaged Hamiltonian known as "secular". 
The Hamiltonian $\mathcal{H}$ is averaged over the fast dynamical timescale, the orbital periods.
In presence of MMR, averaging is not possible since some combinations of mean longitudes (describing the motion over the orbits) are no longer fast with respect to the secular dynamics.
In particular, the conservation of the semi-major axis no longer holds in this configuration.

The averaging approach leads to the averaged Hamiltonian (denoted $\langle\mathcal{H}\rangle$) over the mean longitudes $\lambda_k=M_k + \varpi_k$, ($M_k$ the mean anomaly of body $k$), given as
\begin{equation}
    \langle\mathcal{H}\rangle = \mathcal{H}_0 + \langle\mathcal{H}_1\rangle,
\end{equation}
with $\langle\mathcal{H}_1\rangle=\mathcal{H}_{\mathrm{sec}}$, expressed as
\begin{equation}
    \mathcal{H}_{\mathrm{sec}}=-\sum_{k=1}^N\sum_{l=1}^{k-1}\frac{\mathcal{G}m_km _l}{(2\pi)^2}\int_0^{2\pi}\int_0^{2\pi}\frac{d\lambda_kd\lambda_l}{||\mathbf{r}_k-\mathbf{r}_l||},
    \label{eq:H_sec}
\end{equation}
and where $\mathbf{r}_k$ is the canonical heliocentric coordinate \citep{Poincare_1905} of body $k$.

Since the effect of General Relativity (GR) is non-negligible in the Solar System \citep{Laskar_2008}, we  consider the secular non-constant contributions of GR to the Hamiltonian of the system as derived in \citet{Saha_1992}, \citet{Mogavero_2017} or \citet{Tremaine_2023}.

\subsection{Angular momentum deficit (AMD)}

The conservation of the angular momentum $\mathbf{L}$ of the system leads to a natural choice for the orientation of the frame, such that $\mathbf{L}$ is aligned with the $z$-axis. 
With this choice the norm $L$ of $\mathbf{L}$ becomes 
\begin{equation}
	L=\sum_{k=1}^N\Lambda_k\sqrt{1-e_k^2}\cos(i_k).
\end{equation}
In the secular framework, as shown by Laplace, the semi-major axes of planets are constants.
The conservation of semi-major axes leads to the conservation of the Angular Momentum Deficit.
This quantity was first introduced by \citet{Laplace_1784} and then later used in \citet{Laskar_1997, Laskar_2000} and in \cite{Laskar_2017} to assess the stability of planetary systems. 
The \AMD is defined as the difference between the angular momentum that planets would have on coplanar, circular orbits and the norm of $\mathbf{L}$, i.e.
\begin{equation}
	\AMD=\sum_{k=1}^N\Lambda_k\left(1-\sqrt{1-e_k^2}\cos(i_k)\right).
\end{equation}
We can express the AMD as a sum of the eccentricity \AMD (\AMDe) and inclination AMD (\AMDi) using Poincar\'e's variables $x$ and $y$ defined in Eq. \eqref{eq:x_y_var}.
We then have
\begin{equation}
\AMD = \underbrace{\sum_{k=1}^N x_k\bar{x}_k}_{\AMDe} + \underbrace{\sum_{k=1}^N y_k\bar{y}_k}_{\AMDi} \simeq \sum_{k=1}^N \frac{\Lambda_k}{2} e_k^2 + \sum_{k=1}^N \frac{\Lambda_k}{2} i_k^2.
\label{eq:AMD}
\end{equation}
The \AMD can thus be seen as a weighted sum of the squared eccentricities and inclination of a system.
It measures the excitation of the system, in other words, the non linearity of the orbital dynamics.\\

To determine the stability of a planetary system, we can compare the system's \AMD to a certain critical value beyond which the system does not satisfy the associated stability criterion.
We shall use here, as criterion, the upper limit of \AMD that still satisfies the Hill stability criterion and which is given by \citet{Petit_2018} for a pair of planets, which we denote $\AMD_{\mathrm{H}}^{\mathrm{(k,k+1)}}$.

In the worst dynamical scenario, the entire \AMD of a system is absorbed and contained within one pair $(k,k+1)$ of planets.
If this is the case, the system might become dynamically unstable if this amount of \AMD is such that the Hill critical \AMD for this pair, $\AMD_{\mathrm{H}}^{\mathrm{(k,k+1)}}$, is smaller than the \AMD of the system (fully contained in the pair).
In other words the case ${\AMD_{\mathrm{H}}^{\mathrm{(k,k+1)}}<\AMD}$ implies that the system can technically become unstable.
For simplicity we denote as $\mathrm{AMD_H}$, the minimal pair Hill critical AMD. We call critical AMD coefficient the ratio $\AMD_0/\AMDH$.
We shall thus use this criterion as a proxy for the typical limiting AMD ensuring stability in systems.
This criterion is stricter than the collision condition for secular dynamics that was developed in \citet{Laskar_2017}. 
\subsection{Laplace-Lagrange solution}\label{Laplace-Lagrange}
There is no analytical expression in Poincar\'e  variables for the averaged secular Hamiltonian.
However, it is possible to study analytically the truncated expansion in power series of $(x_k, y_k)$.
In particular, the development of the solution of degree two was proposed by \citet{Lagrange_1778} for the inclinations and \citet{Laplace_1775} for the eccentricities\footnote{Laplace was, in fact, inspired by Lagrange's earlier work, which had been submitted to the French Acad\'emie before 1775 but was only published in 1778.}.
For small eccentricities and inclinations, the higher degree terms of the secular Hamiltonian (Eq. \ref{eq:H_sec}) is a perturbation to the integrable Laplace-Lagrange problem.
The LL Hamiltonian ($\mathcal{H}_{\mathrm{LL}}$), is the truncation of the secular Hamiltonian at degree 2 in Poincar\'e’s variables and can be expressed as \citep[see e.g.][]{Laskar_1995}
\begin{equation}
	\mathcal{H}_{\mathrm{LL}}=\bar{\mathbf{x}}^\mathrm{T}\cdot\mathbf{Q_x}\cdot\mathbf{x}+\bar{\mathbf{y}}^\mathrm{T}\cdot \mathbf{Q}_\mathbf{y}\cdot \mathbf{y},
\end{equation}
where we have defined the column vectors $\mathbf{x}=(x_1,...,x_N)^\mathrm{T}$ and \smash{$\mathbf{y}=(y_1,...,y_N)^\mathrm{T}$}, and where $\mathbf{Q_x}$ and $\mathbf{Q_y}$ are both $N\times N$ symmetric matrixes depending solely on the masses and semi-major axes of the bodies.
The coefficients of these two matrices are given by \citep[see e.g.][]{Petit_2019}
\begin{equation}
\mathbf{Q_x}^{k,l} = 
\begin{cases}
  -\dfrac{\mathcal{G}m_k m_l C_1(\alpha_{k,l})}{a_l\sqrt{\Lambda_k\Lambda_l}}, & \text{for } k \ne l, \\
  \sum\limits_{j \ne l} -\dfrac{\mathcal{G} m_k m_j C_2(\alpha_{k,j})}{a_j\Lambda_k}, & \text{for } k = l,
\end{cases}
\label{eq:MatrixQx}
\end{equation}
and
\begin{equation}
\mathbf{Q_y}^{k,l} = 
\begin{cases}
  -\dfrac{\mathcal{G}m_k m_l C_2(\alpha_{k,l})}{a_l\sqrt{\Lambda_k\Lambda_l}}, & \text{for } k \ne l, \\
  \sum\limits_{j \ne l} \dfrac{\mathcal{G} m_k m_j C_2(\alpha_{k,j})}{a_j\Lambda_k}, & \text{for } k = l,
\end{cases}
\label{eq:MatrixQy}
\end{equation}
with $\alpha_{k,l}=a_k/a_l$ and
\begin{equation}
	C_1({\alpha})=-\frac{\alpha}{4}	b_{3/2}^{(2)}(\alpha), \quad C_2(\alpha)=\frac{1}{4}\alpha b_{3/2}^{(1)}(\alpha),
\end{equation}
where $b_{s}^{(k)}(\alpha)$ are the usual Laplace coefficients \citep[see e.g.][]{Laskar_1995,Tremaine_2023}.
These coefficients are such that the matrices in Eqs. \eqref{eq:MatrixQx} and \eqref{eq:MatrixQy} are symmetric.

Finally, in the LL formalism, the equations of motion are
\begin{equation}
	\frac{\mathrm{d}\mathbf{x}}{\mathrm{d}t}=-\iota\frac{\partial\mathcal{H}_{\mathrm{LL}}}{\partial \bar{\mathbf{x}}}=-\iota\mathbf{Q_x}\cdot \mathbf{x}, \quad \frac{\mathrm{d}\mathbf{y}}{\mathrm{d}t}=-\iota\frac{\partial\mathcal{H}_{\mathrm{LL}}}{\partial \bar{\mathbf{y}}}=-\iota\mathbf{Q_y}\cdot \mathbf{y}.
	\label{eq:eq_motion_LL}
\end{equation}
These define matrices with constant coefficients.
Since the matrices are real and symmetric, they can be diagonalized through orthogonal matrices $\mathrm{\mathbf{O_x}}$ and $\mathrm{\mathbf{O_y}}$, i.e. we have $\mathbf{Q_x}=\mathbf{O_x}\cdot \mathbf{D_x}\cdot \mathbf{O_x}^\mathrm{T}$ and $\mathbf{Q_y}=\mathbf{O_y}\cdot \mathbf{D_y}\cdot \mathbf{O_y}^\mathrm{T}$.
Thus, the elements of the diagonal matrices $\mathbf{D_x}$ and $\mathbf{D_y}$ are the eigenvalues of $\mathbf{Q_x}$ and $\mathbf{Q_y}$ respectively and the columns of $\mathbf{O_x}$ and $\mathbf{O_y}$ are their eigenvectors.
The solution to Eq. \eqref{eq:eq_motion_LL} is a sum of periodic terms whose frequencies are given by the eigenvalues.

The frequencies associated to the $x$ Poincar\'e  coordinates are denoted $g_j$ and the ones associated to $y$ are denoted $s_j$.
Because of the conservation of angular momentum, one of the $s_j$ frequencies is zero\footnote{$s_5$ in the case of the Solar System.}.
In fact this is a consequence of the rotation invariance of the system \citep[see e.g.][]{Morbidelli_2002}.
Using the previous developments, new canonical variables can be introduced.
We call the complex variables $\mathbf{x}^\ast$ and $\mathbf{y}^\ast$ the proper modes of the $\mathbf{x}$ and $\mathbf{y}$ Poincar\'e  variables respectively.
We have
\begin{equation}
\mathbf{x}^\ast=\mathbf{O_x}^{\mathrm{T}}\cdot \mathbf{x}, \quad \mathbf{y}^\ast=\mathbf{O_y}^{\mathrm{T}}\cdot \mathbf{y}.
\end{equation}
These transformations are in fact a simple rotation and thus correspond to the canonical transformation $(\mathbf{x},-\iota\bar{\mathbf{x}})\,\rightarrow(\mathbf{x}^\ast,-\iota\bar{\mathbf{x}}^\ast)$,  $(\mathbf{y},-\iota\bar{\mathbf{y}})\,\rightarrow(\mathbf{y}^\ast,-\iota\bar{\mathbf{y}}^\ast)$.
Consequently we can write
\begin{equation}
\begin{aligned}
     x_{\mathrm{LL},k}(t)&=\sum_j O_{x, k,j}x_j^\ast(0) \mathrm{E}^{\iota g_j t},\\
     y_{\mathrm{LL},k}(t)&=\sum_j O_{y, k,j}y_j^\ast(0) \mathrm{E}^{\iota s_j t}.
     \label{eq:Poincare_reconstructed}
\end{aligned}
\end{equation}
The proper modes were first used by Laplace and Lagrange and their use was then extensively (and more recently) described in \citet{Laskar_1990}.
The main idea is that the proper modes are coordinates which are more suitable for the analysis of the solution of the Hamiltonian.
Indeed the $\mathbf{x}$ and $\mathbf{y}$ temporal evolutions are less practical because they can be dominated by the quasi-periodic motion of the LL dynamics, whereas the proper modes are not supposed to vary significantly if the system behaves very regularly. 
\subsection{Resonances and AMD components exchanges}\label{sec:Resonances_AMD_exchanges}
The secular Hamiltonian, $\mathcal{H}_{\mathrm{sec}}$, expressed via the action and angle pairs $(\mathbf{I}^\ast, \bm{\theta}^\ast)$ and $(\mathbf{I}'^\ast, \bm{\theta}'^\ast)$, originating from the proper mode transformation
\begin{equation}
    \mathbf{x}^\ast=\sqrt{\mathbf{I}^\ast}\,\mathrm{E}^{-\iota\bm{\theta}^\ast}$ and $\mathbf{y}^\ast=\sqrt{\mathbf{I}'^\ast}\,\mathrm{E}^{-\iota\bm{\theta}'^\ast},
    \label{eq:Action_Angle}
\end{equation}
can be expressed as a Fourier series (see Appendix \ref{appendix:Hsec} for more details).
The latter reads
\begin{equation}
    \mathcal{H}_{\mathrm{sec}}=\sum_{\mathbf{k,k'}}\mathrm{E}^{\iota(\mathbf{k}\cdot\bm{\theta}^{\ast}+\mathbf{k}'\cdot\bm{\theta}'^{\ast})}\prod_{j=1}^N I^\ast_j{}^{\frac{|k_j|}{2}} I'^{\ast}_{j}{}^{\frac{|k'_j|}{2}}(\tilde{h}^\ast{}^{\mathbf{k}, \mathbf{k}'}(\bm{\Lambda}, \mathbf{I}^\ast, \mathbf{I}'^{\ast})),
    \label{eq:H_sec_Fourier}
\end{equation}
with $\mathbf{k}$ and $\mathbf{k'}$ two integer coefficient vectors in $\mathbb{Z}^N$. 
The order of an harmonic in $\mathcal{H}_{\mathrm{sec}}$ is then defined as the 1-norm\footnote{The 1-norm, or taxicab norm, of a vector, is simply the sum of the absolute values of the vector's components.} $\lVert (\mathbf{k}, \mathbf{k}') \rVert_1$.
Note that the d'Alembert rules imply that, due to the invariance by rotation of the system, any harmonic must be such that $\sum_j (k_j+k_j')=0$.
If an harmonic becomes at some point resonant, i.e. $\phi_{\mathrm{res}}$ the argument of the exponential in Eq. \eqref{eq:H_sec_Fourier} is slow, one has $\dot{\phi}_{\mathrm{res}}=\mathbf{k}\cdot\dot{\bm{\theta}}^\ast+\mathbf{k'}\cdot\dot{\bm{\theta}}'^\ast\approx 0$, chaos may arise from overlap with neighboring resonances.
Exchanges between the eccentricity and inclination degrees of freedom can then occur.
However, not all resonant arguments trigger exchanges.
The resonant frequency associated with a given harmonic, is approximated as a linear combinations of the g and s secular frequencies of the system, i.e. $\dot{\phi}_{\mathrm{res}} \approx \mathbf{k}\cdot\mathbf{g} + \mathbf{k'}\cdot\mathbf{s}$.  
Indeed the secular frequencies vectors are dominated by the Laplace-Lagrange dynamics, i.e.
\begin{equation}
    \dot{\bm{\theta}}^\ast=\frac{\partial \mathcal{H}_{\mathrm{sec}}}{\partial \mathbf{I}^\ast} = \mathbf{g} +O(e^2), \quad \dot{\bm{\theta}}'^\ast=\frac{\partial \mathcal{H}_{\mathrm{sec}}}{\partial \mathbf{I}'^\ast} = \mathbf{s} +O(i^2).
\end{equation}
A particular feature of the LL dynamics is that all $g$ frequencies are positive and all $s$ frequencies are negative \citep[see, e.g,][for a proof]{Saillenfest_2019}.
The key idea behind the existence of the sub-reservoirs and thus the quasi-integral nature of \AMDe and \AMDi is that exchanges between these two sub-reservoirs are rare and occur only when specific resonances are excited.
More precisely for a given harmonic, according to Appendix \ref{appendix:Hsec}, the temporal derivatives of the \AMD, \AMDe and \AMDi read
\begin{equation}
\dot{\AMD}\propto\sum_j k_j+k'_j,\quad \dot{\AMDe} \propto  \sum_j k_j, \quad \dot{\AMDi} \propto  \sum_j k'_j.
\label{eq:AMD_dot}
\end{equation}

On the one hand we can see that the first equation shows the conservation of the AMD.
Indeed we have $\dot{\AMD}\propto \sum_j k_j+k'_j$, which, following the d'Alembert rules, is identically $0$, for any harmonic ($\mathbf{k},\mathbf{k'}$).
Consequently no secular resonance breaks the AMD conservation.
The conservation of the AMD is in fact a simple consequence of the d'Alembert rules.
On the other hand, according again to Eq. \eqref{eq:AMD_dot}, only harmonics such that $\sum_j k'_j\neq0$ (resp. $\sum_j k_j\neq0$) can lead to non-conservation of \AMDi (resp. \AMDe) \citep[see also][]{Mogavero_2023}. 
We point out that, due again to the d'Alembert rules, having $\sum_j k'_j\neq0$ implies $\sum_j k_j\neq0$. Therefore studying the conservation of \AMDe and \AMDi is fairly equivalent.

\begin{figure}%
\begin{center}
\includegraphics[width=1.0\linewidth]{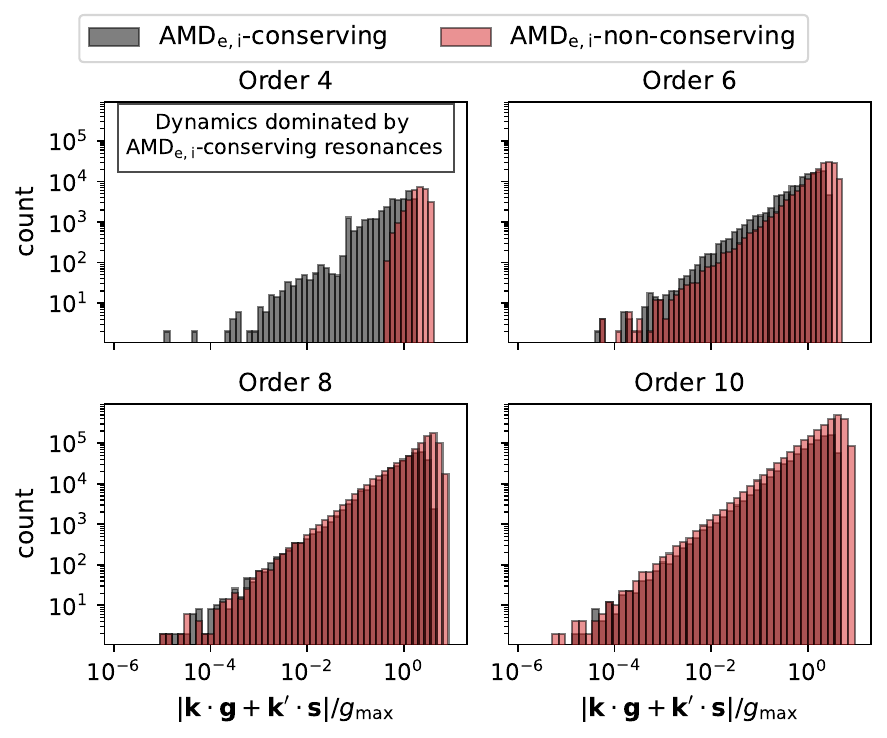}
\caption{Histograms of normalized secular resonance frequencies using the Laplace-Lagrange frequencies of the 250 systems within the \texttt{Peas}$_3$ setup.
Harmonics' strength are compared for each order in the cases allowing and not allowing inclination-eccentricity AMD exchanges.
}
\label{fig:AMDi_conserving_vs_non_conserving_freqs}
\end{center}
\end{figure}

Satisfying the resonance condition $\mathbf{k}\cdot\mathbf{g}+\mathbf{k}'\cdot\mathbf{s}\approx0$ is usually harder with two large contributions canceling out (i.e. $\mathbf{k}\cdot\mathbf{g}\approx-\mathbf{k'}\cdot\mathbf{s}$) than with both terms independently small (i.e. $\mathbf{k}\cdot\mathbf{g}\approx0$ and $\mathbf{k}'\cdot\mathbf{s}\approx0$).
The latter configuration is more readily achieved at low order for harmonics satisfying $\sum_j k_j = \sum_j k'_j =0$, i.e. for $\mathrm{AMD_{i,e}}$-conserving harmonics.
However at high order, the larger components in $\mathbf{k}$ and $\mathbf{k}'$ provides more freedom to achieve either resonant configuration, even with $\sum_j k_j$ and  $\sum_j k'_j\neq0$.
Consequently, resonances allowing exchanges of \AMD between the eccentricity and inclination reservoirs generally appear only at higher order.
We show in Fig. \ref{fig:AMDi_conserving_vs_non_conserving_freqs} all possible values of $\dot{\phi}_{\mathrm{res}}$ for the \texttt{Peas}$_3$ setup (see Table. \ref{tab:Simulation_params}) by computing \citep[using the celmech package, see][]{Hadden_2022} the LL $\mathbf{g}$ and $\mathbf{s}$ secular frequencies for all systems within the setup and considering different orders of harmonics.
This figure highlights the idea that as the order increases, the number of near resonant tuples (i.e. for which $|\mathbf{k}\cdot\mathbf{g}+\mathbf{k}'\cdot\mathbf{s}|\ll g_{\mathrm{max}}$), increases.
We note that order 4 does not allow active (i.e. $\approx 0$) and non-$\mathrm{AMD_{i,e}}$-conserving resonances. On the other hand higher order harmonics do allow active resonances that are non-$\mathrm{AMD_{i,e}}$-conserving.
This suggests that the resonance network favors the quasi-conservation of \AMDe and \AMDi at low order.
Moreover higher-order harmonics are typically weaker, so the associated resonances generally have smaller widths and shorter effective lifetimes.
The extent to which \AMDe and \AMDi remain conserved therefore depends on how strongly the dynamics are influenced by higher-order harmonics.

\section{Numerical  simulations setups}\label{sec:Numerical}
Given that this study relies on numerical simulations of multiplanetary systems, we describe here the underlying principles and accuracy of the numerical codes used.
We used both a secular dynamics code and the N-body integrator REBOUND \citep{Rein_2012} to perform our study.
We give here a description and integration comparison of these two codes.
\subsection{N-body integrations}
The REBOUND package is used to integrate systems using two improved implementations of the unbiased symplectic Wisdom-Holman integrator \citep{Wisdom_1991}. We use the integrators WHFast \citep{Rein_2015} with the $11^{\mathrm{th}}$ order symplectic corrector for the \texttt{Peas} and \texttt{Far} series (100 Myr integrations, see Table. \ref{tab:Simulation_params}) and WHFast512 \citep{Dagli_2026} for the \texttt{OutPeas} and \texttt{8Peas} series (300 and 600 Myr integrations respectively, see Table. \ref{tab:Simulation_params}), and $17^{\mathrm{th}}$ order symplectic corrector.
Following \citet{Wisdom_2015}, we set the integrators with a timestep equal to 1/20 of the minimum perihelion passage timescale.
The contribution from general relativity is taken into account using the \texttt{gr\_potential} effect from the REBOUNDx package \citep{Tamayo_2020}.
The initial mean longitudes are randomly drawn. 
We enforce the simulations to stop when the distance between two planets is inferior to twice the Hill radius of the second planet of the system (i.e. $r_{\mathrm{Hill}}\approx a_2\sqrt{m_2/3m_0}$).
In such case we consider the system to be unstable.

\subsection{Secular integrations}\label{sec:presentation_of_Gausscode}
Integrating a secular Hamiltonian allows one to study the dynamical evolution of a planetary system over very long timescales, by reproducing the longterm behavior of N-body integrations, as long as the system is sufficiently far from MMRs, as is the case in our study.
The trajectories generated by the secular Hamiltonian $\mathcal{H}_{\mathrm{sec}}$ in Eq. \eqref{eq:H_sec} can be integrated numerically in a very efficient way using the so-called Gauss method \citep{Gauss_1818,Bour_1855,Hill_1882}. Here we employ the implementation developed in \citep{Mogavero_2021} for the study of the inner planets in the Solar System, which we generalized to any planetary system configuration and which we refer to as “Gausscode”. 
All Gausscode integrations are performed using an Adams PECE method of order 12 with a timestep set to $1/200$ the shortest secular period among the system's inclination and eccentricity LL modes.
With such a choice we typically have $\Delta t_{\mathrm{REBOUND}}/\Delta t_{\mathrm{Gausscode}}\sim 10^{-4}-10^{-5}$.
A Gausscode integration is stopped when a secular collision happens, either a star-planet or a planet-planet one. 
The second case takes place when there is an intersection of the instantaneous Keplerian orbits of a pair of planets \citep{Laskar_1994, Mogavero_2021}.
This cannot be seen as an actual physical collision between the bodies, but is simply a geometrical intersection of orbital paths.

\subsection{Simulated systems}
We consider four series of systems, namely \texttt{Peas}, \texttt{OutPeas}, \texttt{8Peas} and \texttt{Far}, each made of 4-planet systems, except for \texttt{8Peas} that are 8-planet systems. Each planet has a mass set to $10^{-5}\,\Ms$ (or approximately $3.3\,\Me$).
We thereby ensure to simulate a system of super-Earths which is typical of many exoplanetary systems.
The \texttt{Peas}, \texttt{OutPeas} and \texttt{8Peas} series have the inner planet's semi-major axis set to 0.1 AU and the other planet's semi-major axes are computed by drawing 3 period ratios following a uniform distribution between 1.77 and 1.93.
We choose similar period ratios to remove from our results the strong dependence in period ratio of the secular frequencies.
These period ratio values ensure to be outside regions where the dynamics is driven by mean motion resonances \citep[see e.g.][]{Tamayo_2021}.
The typical semi-major axes thus range between 0.1 AU and 0.37 AU. We consider in this case a typical close-by super-Earth system. 
These series are consistent with the "peas in a pod pattern" \citep{Weiss_2018}, hence the "Peas" in their names.
The \texttt{OutPeas} systems have the total \AMD only distributed between the two outermost planets whereas the other sets are such that the \AMD is roughly equipartitioned between all planets.
The REBOUND integrations of the \texttt{Peas}, \texttt{OutPeas} and \texttt{8Peas} series (and Gausscode integrations for \texttt{Peas}) do include post-Newtonian corrections as it is non-negligible for such close planets.
The last series of systems is called "\texttt{Far}", we consider a constant period ratio of adjacent pairs of 3.15, with the inner planet's semi-major axis set to $1\,\,\mathrm{AU}$.
The semi-major axes considered are then $1\,\,\mathrm{AU}$, $2.15\,\,\mathrm{AU}$, $4.62\,\,\mathrm{AU}$, $9.92\,\,\mathrm{AU}$. 
The REBOUND integrations of the \texttt{Far} systems do not include post-Newtonian correction as it's contribution to the dynamics is negligible for such systems.
We consider 7 different setups of these \texttt{Peas}, "\texttt{Far}" series and 2 setups of the \texttt{OutPeas} series. Dynamically, the only difference between the setups of a given series is the initial total \AMD ($\AMD_0$) considered and thus the range of initial eccentricity and inclinations.
Within each setup, all systems initially have the same amount of \AMD, and are such that $\AMDe=\AMDi=\AMD/2$, i.e. we consider equipartition between the eccentricity and inclination degrees of freedom.
This is motivated by the work of \citet{Xie_2016} where it was found that typical systems from the Kepler data were such that the average $e$ and $i$ were of the same order of magnitude.
Moreover, the Solar System 8 planets are approximately in this equipartition state.
We detail the way to initialize systems given a value of eccentricity and inclination \AMD in Appendix \ref{appendix:initialization_systems}.
We summarize the simulations initial parameters and survival rates of setups considered in Table \ref{tab:Simulation_params}.
\section{Results}\label{sec:Results}
We present here the results of the different numerical simulations of the \texttt{Peas} and \texttt{Far} series performed via both REBOUND and Gausscode in the scope of the study of the conservation of the \AMDe, \AMDi and proper modes.
We will use, as a proxy for the conservation of these quantities, the standard deviation (std), computed over the time evolution of the quantities.
We will also always consider the \AMD-normalized stds, allowing for meaningful comparisons across quantities and dynamical regimes.
\subsection{Variability under the ergodicity hypothesis}
Under the ergodic hypothesis, there is no conservation of \AMDe and \AMDi.
More precisely, systems can be, at any instant, in a configuration where the \AMD is arbitrarily distributed between \AMDe and \AMDi.
Consequently the inclinations and eccentricities of planets can be such that \AMDe and \AMDi can take any value from 0 to the total AMD (but still ensuring that $\AMDe + \AMDi = \AMD$).
This ergodic hypothesis essentially allows us to probe how much standard deviations we should expect if exchanges between \AMDe and \AMDi were not restricted along the evolution of the system.
In order to assess this upper limit, we drew one million initial conditions for the \texttt{Far}$_2$ system setup.
We detail the process to compute the standard deviations under the ergodicity hypothesis in Appendix \ref{appendix:ergodicity}.
We stress that the choice of setup considered is fully arbitrary. With four planets, the dynamics is equivalent to a uniform diffusion over a hypersphere $\mathbb{S}^{15}$ of radius $\sqrt{\AMD}$.
The value of the \AMD considered is irrelevant, only the dimension of the problem matters, which is only a function of the number of planets.
Therefore the result is independent of the critical AMD coefficient.
Using these initial conditions we found an AMD-normalized standard deviation of about $0.11$ and $0.17$ for the squared proper modes and the \AMDi (and \AMDe) respectively.
We used these values of variability in the ergodic dynamical hypothesis as a typical upper limit for the conservation of the modes and of the \AMD components. 
Under the ergodic hypothesis the time average of a given quantity is equivalent to its spatial average.
Therefore, if the phase space is not restrained by the quasi-integrals, i.e. if the quasi-integrals do not exist, we expect the temporal standard deviations of \AMDe and \AMDi to be of the order of $0.17$ after normalization.
Since the proper modes have, a priori, no reason to be conserved, we expect their standard deviations to be of the same order of magnitude than the one in the ergodic hypothesis (i.e. 0.11).

\begin{table*}
\caption{
Summary of the simulation setups
}
    \centering
\resizebox{\textwidth}{!}{%
{\small
\begin{tabular}{c| c c c c c c c c c c c c}
\hline\hline 
Setup & $N$ & $m$\,\,[\Ms]  & Period ratio &  $a_{\mathrm{in}}\,$[AU] & $N_{\mathrm{systems}}$ & IT$_{\mathrm{REBOUND}}$ [Myr] &  IT$_{\mathrm{Gausscode}}$ [Myr] & SR$_{\mathrm{REBOUND}}$ [\%] & SR$_{\mathrm{Gausscode}}$ [\%] & $\frac{\AMDe}{\AMDi}$ & median e$_0$ & $\AMD_0/\mathrm{median (AMD_H)}$ \\
\hline
\texttt{Peas}$_1$ & \multirow{7}{*}{4} & \multirow{7}{*}{$10^{-5}$} & \multirow{7}{*}{$\sim \mathcal{U}$(1.77,1.93)} & \multirow{7}{*}{0.1} &\multirow{7}{*}{250} & \multirow{7}{*}{100} & \multirow{7}{*}{1000} &   \multirow{4}{*}{100} & \multirow{5}{*}{100} & \multirow{7}{*}{1} & 0.006  &  $0.009$\\
\texttt{Peas}$_2$ & & & & & & & & & & & 0.01 &$0.021$\\
\texttt{Peas}$_3$ & & & & & & & & & & & 0.028 &$0.148$\\
\texttt{Peas}$_4$ & & & & & & & & & & & 0.046 &  $0.446$\\
\texttt{Peas}$_5$ & & & & & & & & $99.6$& & &0.063 &  $0.746$\\
\texttt{Peas}$_6$ & & & & & & & & $89.2$& 98.4 & &0.075&   $1.111$\\
\texttt{Peas}$_7$ & & & & & & & & $48.4$& 86.8 & &0.090 &  $1.647$\\
\hline
\texttt{OutPeas}$_3$ &\multirow{2}{*}{4} & \multirow{2}{*}{$10^{-5}$} & \multirow{2}{*}{$\sim \mathcal{U}$(1.77,1.93)} & \multirow{2}{*}{0.1}  &\multirow{2}{*}{80} & \multirow{2}{*}{300} & \multirow{2}{*}{\textemdash}& $100$&  \multirow{2}{*}{\textemdash}& \multirow{2}{*}{1} & \multirow{2}{*}{\textemdash}& $0.147$ \\
\texttt{OutPeas}$_6$ & & & & & & & & $82.5$&  & & &  $1.088$\\
\hline
\texttt{8Peas} & 8 & $10^{-5}$& $\sim \mathcal{U}$(1.77,1.93)&0.1 & 80& 600& \textemdash& 46.25& \textemdash & 1 & 0.060 & 2.260 \\
\hline
\texttt{Far}$_1$ & \multirow{7}{*}{4} & \multirow{7}{*}{$10^{-5}$} & \multirow{7}{*}{3.15} & \multirow{7}{*}{1} &  \multirow{7}{*}{250} & \multirow{7}{*}{100} & \multirow{7}{*}{5000}& \multirow{6}{*}{100} & \multirow{3}{*}{100} &\multirow{7}{*}{1} & 0.009 & $0.006$\\
\texttt{Far}$_2$ & & & & & & & & & & &0.046 & $0.151$\\
\texttt{Far}$_3$ & & & & & & & & & & & 0.092 & $0.605$\\
\texttt{Far}$_4$ & & & & & & & & & 98.8 & &0.116 & $0.944$\\
\texttt{Far}$_5$ & & & & & & & & & 97.2 & & 0.135 & $1.360$\\
\texttt{Far}$_6$ & & & & & & & & & 92 & & 0.163 & $1.852$\\
\texttt{Far}$_7$ & & & & & & & & 99.2& 63.2& &0.187  & $2.418$\\
\hline
\end{tabular}
}
}
\tablefoot{Initial conditions (and survival rates "SR", at integration time "IT") from the "\texttt{Peas}", "\texttt{OutPeas}", "\texttt{8Peas}" and "\texttt{Far}" series used to study the conservation of \AMDi and \AMDe under varying initial conditions.Across all runs, the planetary masses are $M_{\mathrm{pl}}=10^{-5}\,\Ms$ and the initial ratio of \AMD between the eccentricity and inclination degrees of freedom, $\AMDe/\AMDi$, was set to 1, ensuring that $50\%$ of AMD originates from the eccentricity degree of freedom and $50\%$ from the inclination degree of freedom.
The "\texttt{Peas}", "\texttt{OutPeas}" and "\texttt{8Peas}" series have the innermost planet's semi-major axis and adjacent pair period ratio set to $0.1\,\,\mathrm{AU}$ and $\sim \mathcal{U}$(1.77,1.93) respectively. On the other hand the "\texttt{Far}" series have the innermost planet's semi-major axis and adjacent pair period ratio set to $1\,\,\mathrm{AU}$ and $3.15$.
All series have the AMD equipartitioned between planets, except \texttt{OutPeas}, where all the AMD is contained within the two outermost planets.
The "\texttt{Far}", "\texttt{Peas}", "\texttt{OutPeas}" series are 4-planet systems and "\texttt{8Peas}" are 8-planet systems. The "\texttt{OutPeas}" and "\texttt{8Peas}" systems have only been integrated via REBOUND.}
\label{tab:Simulation_params}
\end{table*}

\subsection{Distribution of \texorpdfstring{\AMD, \AMDe and \AMDi}{AMD , AMDe and AMDi} standard deviations}
\begin{figure}%
\begin{center}
\includegraphics[width=1.0\linewidth]{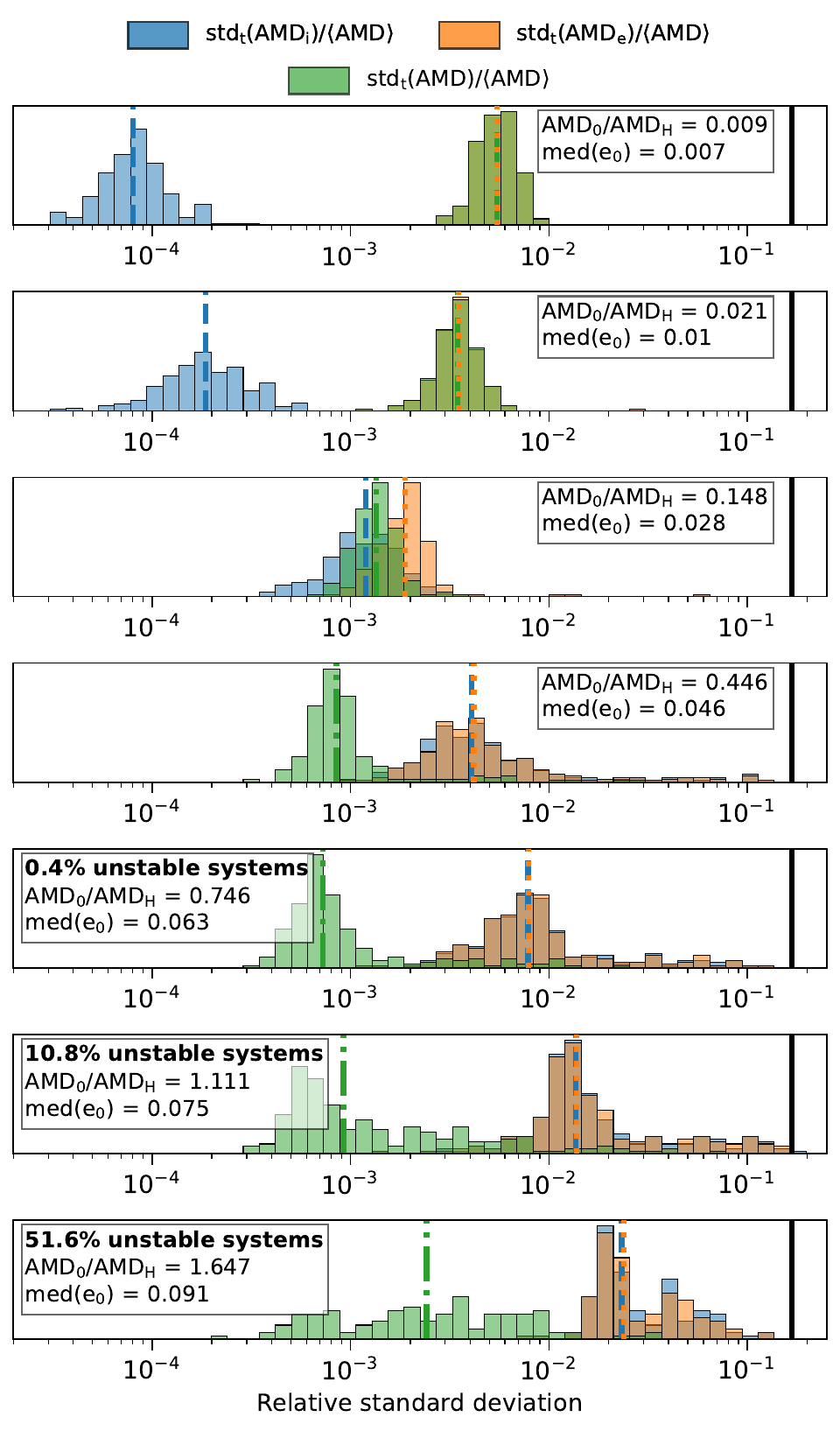}
\caption{Distribution of the standard deviation over the time, of the inclination, eccentricity and total AMD for systems that survived to 100 Myr of the N-body REBOUND integrations.
Each panel corresponds to a given \texttt{Peas} setup, i.e. a given value of initial total \AMD ($\AMD_0$).
Within each panel, we compute the median, across the distribution of simulated systems, of the standard deviations of each quantity.
These medians are shown as dashed, dotted and dash-dotted lines.
The black vertical line at $0.17$ relative standard deviation is the normalized standard deviation of the inclination and eccentricity AMD in the ergodic hypothesis.}\label{fig:AMD_Histograms_No_symmetry}
\end{center}
\end{figure}

We first consider the 250 systems integrated with REBOUND for each setup in the \texttt{Peas} set.
For each system we compute the standard deviation, over the time, of the \AMDi, \AMDe and total \AMD\footnote{We remind that the \AMD is an integral of motion only provided that the planetary semi-major axes remain constant.} as a way to measure the chaotic diffusion of the system variables.
The resulting distributions of standard deviations are shown in Fig. \ref{fig:AMD_Histograms_No_symmetry}. We see that, for stable systems, $\AMD_{\mathrm{e,i}}$ are quasi-integrals of motion.
Indeed, the typical value for the relative standard deviation is only up to the order of the percent.
For very low initial \AMD values, i.e. the \texttt{Peas}$_1$ and \texttt{Peas}$_2$ setups, the \AMD standard deviations are non-negligible and are in fact driven by \AMDe variations.
The \AMDi is on the other hand very well conserved.
Our initial conditions are such that the eccentricities at $t=0$ are below the floor of eccentricity resulting from planet-planet interactions.
When simulating the evolution of these systems, planets tend to very quickly get their eccentricities excited, and thus change their \AMDe (and \AMD).
This phenomenon is not visible for inclinations as its amplitude is negligible. The full perturbation Hamiltonian $\mathcal{H}_1$ is indeed of order $\mathcal{O}(i^2)$ in inclinations whereas it is of order $\mathcal{O}(e)$ in eccentricities.
Consequently, because eccentricities and inclinations are decoupled at very low AMD value, the distributions of both \AMD and \AMDe are approximately the same and \AMDi is very well conserved.
We also notice in Fig. \ref{fig:AMD_Histograms_No_symmetry}  that, as the \AMD, increases, the distribution of its std shifts to lower values and decouples from the \AMDe std distribution.
The std of \AMD then reaches some kind of std range equilibrium and the \AMDe and \AMDi stds distributions become very similar and shift towards increasingly larger values.
Distributions however lose their lognormal-like shape.
We notice, for the \texttt{Peas}$_4$ to \texttt{Peas}$_7$ setups, a tail towards higher standard deviations.
We interpret this as a consequence of an active resonance network in these systems \citep[see e.g.][]{Mogavero_2021}. 
Indeed in the \texttt{Peas} setups, all systems have different planet pairs' period ratios, and thus different sets of $\mathbf{g}$ and $\mathbf{s}$ secular frequencies.
Some systems may be initialized closer to active resonances, and suffer from important variations, hence a tail. 
For such systems, the quasi-integrals do not exist.
For physically stable systems ($\AMD_0/\AMDH<1$) the number of systems within the tail is however small.
Finally, we notice that the standard deviation computed under the ergodicity hypothesis constitutes an almost perfect upper limit for the \AMDe and \AMDi standard deviations.
Due to the stochasticity of the dynamics, it is however not forbidden and even possible for systems to have a standard deviation higher than the one computed assuming ergodicity.

\subsection{Distribution of proper modes standard deviations}
\begin{figure}%
\begin{center}
\includegraphics[width=1.0\linewidth]{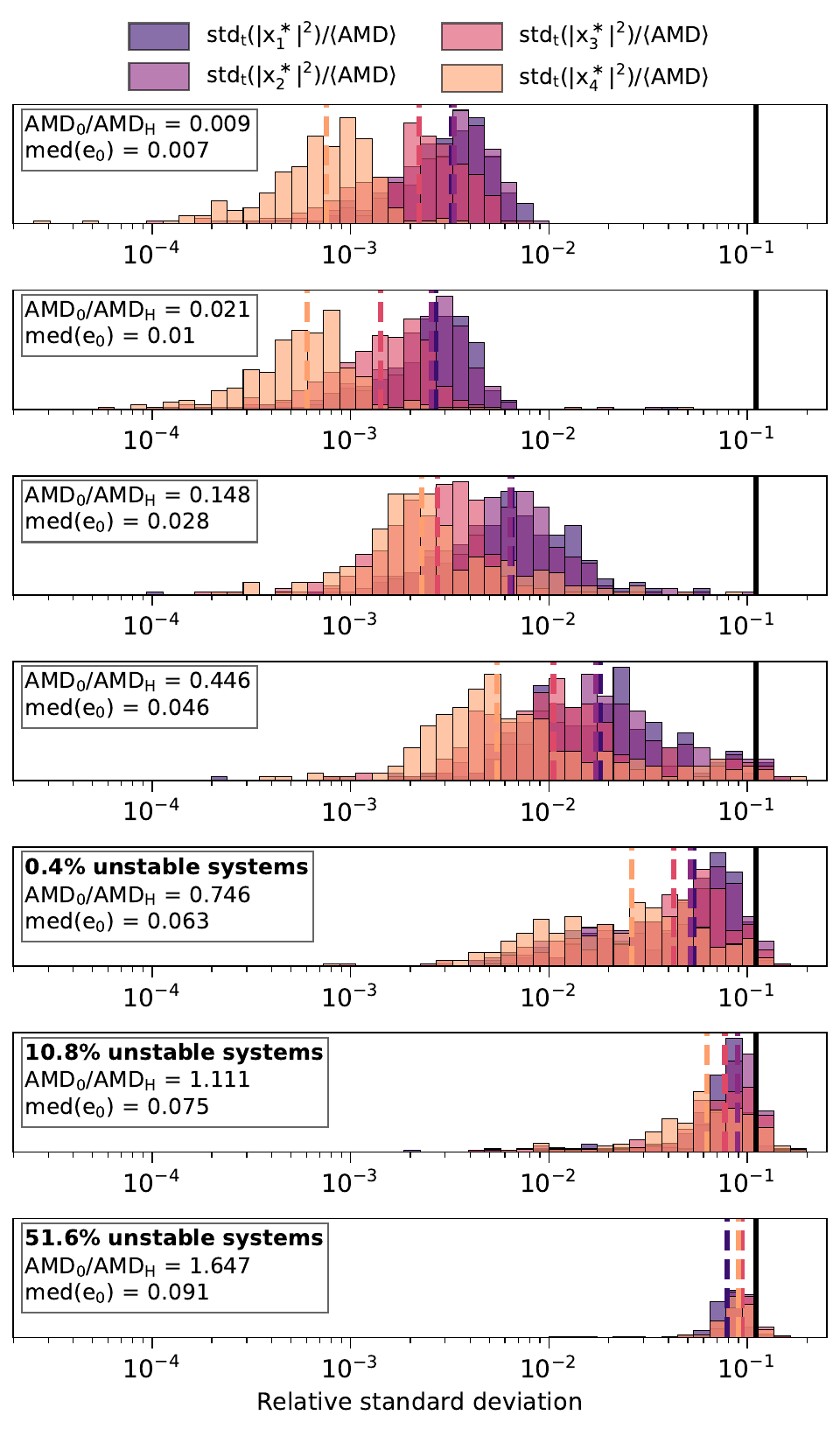}
\caption{Distribution of the standard deviation over the time of the eccentricity proper modes of the systems, for systems that survived to 100 Myr of the N-body REBOUND integrations.
Each panel corresponds to a given value of total initial AMD.
The medians of the 4 distributions are shown as dashed lines for each setup.
The black vertical line at $0.11$ relative standard deviation is the normalized standard deviation of the eccentricity modes in the ergodic hypothesis.}\label{fig:e_modes_Histograms_No_symmetry}
\end{center}
\end{figure}

We now perform the same process but for the eccentricity proper modes, i.e. for each systems within the different \texttt{Peas} setups, we compute the std, over the time dimension, of the squared amplitudes of the modes.
The resulting distributions of standard deviations are shown in Fig. \ref{fig:e_modes_Histograms_No_symmetry}.
For low-AMD systems, the $x_3^\ast$ and $x_4^\ast$ modes, that are "associated" with the two outermost planets, tend to show lower variability.
Given the conservation level of the modes for the $\texttt{Peas}_1$ to $\texttt{Peas}_4$ sets, it appears the systems are almost quasi-periodic.

As the \AMD increases and approaches the stability limit, the distributions of modes get closer to one another and the mode diffusion saturates, i.e. the bulks of the distributions shift to the ergodicity limit.
For Hill stable systems, the distributions are much wider than the ones of \AMDe (and \AMDi).
These features highlight the quasi-integrals nature of both \AMDe and \AMDi.
What is even more striking is that the \AMDe is the sum of the eccentricity modes, therefore given their variations we would expect the \AMDe to be much less conserved (i.e. with a conservation level typically 10 times worse).
The variations of the modes are due to planet-planet interactions and resonances, which, for the most part, are not \AMDe and \AMDi breaking (see Sects. \ref{sec:phenomenology} and \ref{sec:Resonances_AMD_exchanges}).
The inclination proper modes are not shown but exhibit the same overall behavior, except at very low \AMD, where they tend to be even more conserved than the eccentricity modes.
\subsection{Scaling analysis}\label{sec:All_together}
\begin{figure}%
\begin{center}
\includegraphics[width=1.0\linewidth]{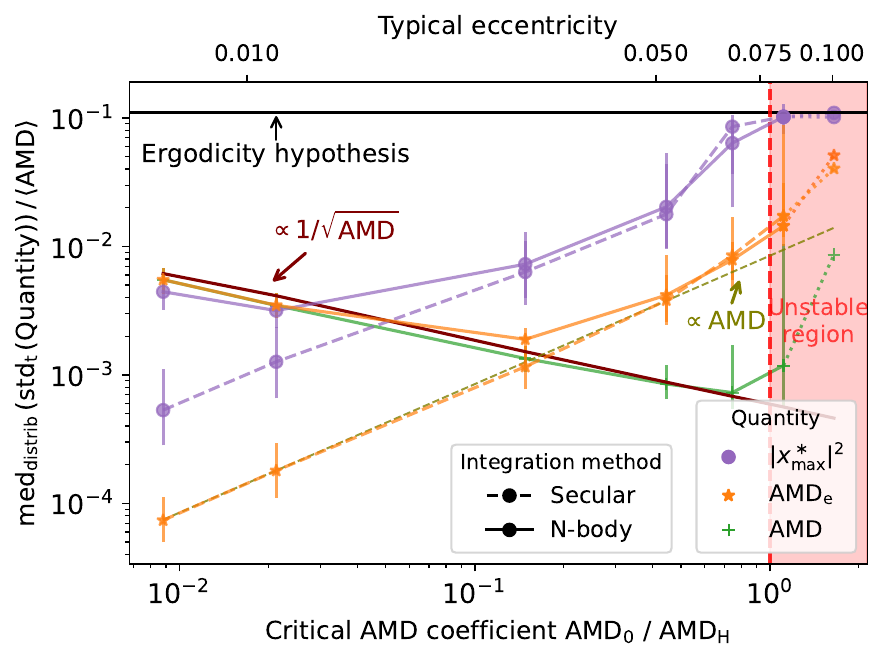}
\includegraphics[width=1.0\linewidth]{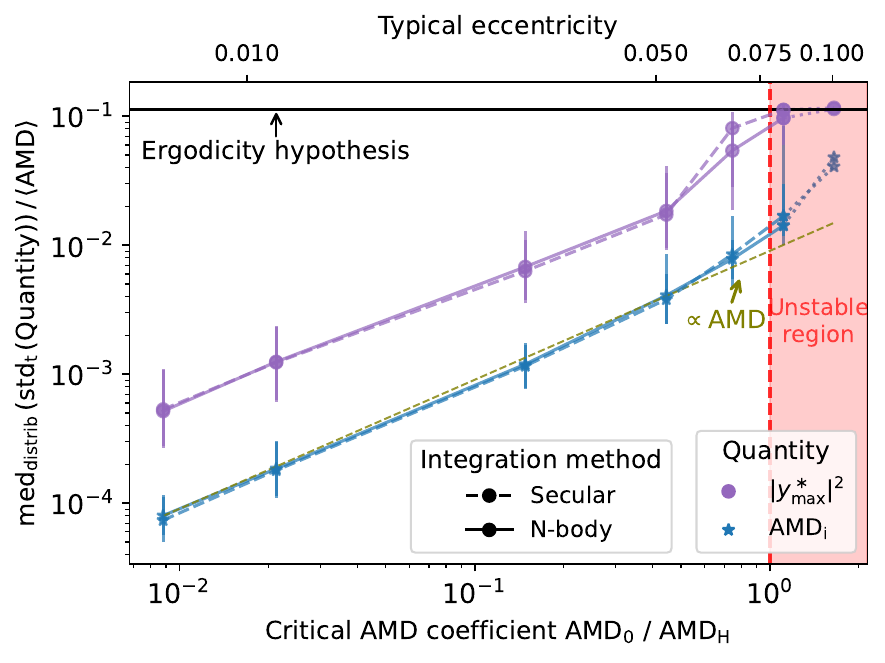}
\caption{Top (resp. bottom) : Median, over the distribution, of the temporal standard deviations of \AMDe (resp. \AMDi) and median, over the distribution of the maximal (over modes) temporal standard deviation of the squared modulus of the x (resp. y) proper modes for both REBOUND and Gausscode simulations as a function of the ratio of the initial total \AMD and the critical Hill \AMD.
The upper axes show the corresponding typical eccentricity.
The red region shows systems initially Hill unstable.
We represent the upper limit for the standard deviation given by the ergodic hypothesis.
Error bars spread the $15^{\mathrm{th}}$ and $85^{\mathrm{th}}$ percentiles of the data.
The error bars for data points of the \texttt{Peas}$_7$ systems are not shown due to the high number of systems colliding during the integrations. 
The trend of Eq.~\eqref{eq:AMD_trend} is shown as a maroon line (Top).
The green dashed line in the top and bottom figures are respectively fits of the \AMDe and \AMDi normalized std over the \texttt{Peas}$_1$ to \texttt{Peas}$_4$ Gausscode simulations. The typical eccentricities are computed based on the amount of \AMDe using the approximation in Eq. \eqref{eq:AMD}.}

\label{fig:Global_No_symmetry}
\end{center}
\end{figure}

We observe, from individual simulation sets, a trend in the conservation levels of the quasi-integrals and of the modes as a function of the total AMD.
In order to obtain a more global view of the results of all simulations within the \texttt{Peas} set, we show in Fig. \ref{fig:Global_No_symmetry} the median over the 250 integrations along with the $15^{\mathrm{th}}$ and $85^{\mathrm{th}}$ percentiles of the time std distributions of \AMD, \AMDe, \AMDi and proper modes with largest variability, for all \texttt{Peas} setups (note that the medians were already shown in Figs. \ref{fig:AMD_Histograms_No_symmetry} and \ref{fig:e_modes_Histograms_No_symmetry}). %
\subsubsection{AMD, AMD components and proper modes variability}
Figure \ref{fig:Global_No_symmetry} suggests that the secular dynamics of compact systems lies between two different dynamical regimes. 
At very low \AMD, the non secular effects naturally excite the eccentricities up to values in the range of $10^{-4}$, which gives a lower bound to the standard deviations of the \AMDe and \AMD.
On the other hand, at higher \AMD, near orbit crossing configurations, the standard deviation distributions of modes are consistent with the ergodicity limit standard deviation.
In contrast with eccentricities, inclinations are practically not excited by non-secular effects at very low \AMD. Therefore the \AMDi and inclination proper modes are very well conserved at low \AMD and the two integration methods give very close results across the entire range of stable systems.
Following what was presented in Sect. \ref{sec:phenomenology}, we can say that, as the \AMD increases, we tend to have more and more non-$\AMD_\mathrm{e,i}$ conserving harmonics that become resonant. In other words some harmonics of order superior to 4 start to be resonant. 
Nevertheless, the level of conservation of \AMDe and \AMDi is still quite striking, the stds being below or of the order of the percent up to the \texttt{Peas}$_5$ setup.
As shown in the previous section, the proper modes, whether of eccentricity or inclinations, are globally not as well conserved as \AMDe and \AMDi.
The squared modulus of proper modes always show larger variations (by a factor of a few) than \AMDi and \AMDe (except at very low \AMD where they are of the same order of magnitude in eccentricities).
Due to a diffusion process, modes vary more and more as the total AMD increases.
Systems can very well be excited by a resonance that does not affect the \AMD components but will affect the modes, thus diffusion in \AMDe and \AMDi is not as strong.
In the end, besides the regime where it is ergodic, the mode amplitudes are not significantly changing (up to the percent level for Hill stable systems).
Additionally, proper modes are only conserved quantities of the LL problem, the fact that they are that much conserved suggests that the dynamics are at most weakly chaotic.
It may still be chaotic as the phases are not modelled here.

\subsubsection{\texorpdfstring{\AMD and $\mathrm{AMD_{e,i}}$}{AMD, AMDe, AMDi} scaling analysis}

We observe in Fig. \ref{fig:Global_No_symmetry} that, at low enough \AMD, the variations of \AMD and \AMDe follow power-law trends.    
More precisely we have the normalized std of AMD scaling as $\AMD^{-1/2}$ and the \AMD-normalized std of \AMDe scaling as \AMD.
We propose here a semi-analytical toy-model to explain the observed trends.
Starting with the \AMD trend, we assume that the dynamics of the proper modes can be modeled as
\begin{equation}
\mathbf{x}^\ast=\mathbf{x}^\ast_0+\mathbf{x}_{\mathrm{kick}}, \quad \mathbf{y}^\ast=\mathbf{y}^\ast_0,
\end{equation}
with $\mathbf{x}^\ast_0$ and $\mathbf{y}^\ast_0$ constants (the secular dynamics are quasi-periodic) and the kick terms are the short term eccentricity variations \citep{Henon_1986} due to the planet conjunctions.
At first order in eccentricities and inclinations there is no such counterpart for the inclinations.
In a secular setting and in absence of three-body MMR, these kicks cancel out and do not lead to this \AMD diffusion \citep{Petit_2020}, but they nevertheless have a non zero impact on the standard deviation of the osculating variables.
Noting $\sigma^2 = \Var(|x_{\mathrm{kick},k}|)$  and using \citet{Henon_1986},
\begin{equation}
    \sigma=\sqrt{\Lambda}e_{\mathrm{floor}}, \quad e_{\mathrm{floor}}=2.23\frac{2m}{3m_0}(1-\alpha)^{-2},
    \label{eq:sigma_Henon_Petit}
\end{equation}
where $\alpha$ is the ratio of the inner to outer planet semi-major axes.
In the \texttt{Peas} systems, $e_{\mathrm{floor}}\sim 10^{-4}$ and $\sqrt{\Lambda}\sim 5\cdot10^{-3}\,\Ms^{1/2}\,\textrm{AU}\,\textrm{yr}^{-1/2}$, consequently $\sigma$, and the components of $x_{\mathrm{kick}}$, are very small with respect to eccentricities.
We then have, to leading order and using the fact that  $x^\ast_{0,k}$ is a constant and under the assumption that the kicks are independent and identically distributed,
$    \Var({x_k^\ast}^2)=4{x_0^\ast}^2\sigma^2.
$ %
We can finally write
\begin{equation}
\Var(\AMDe) = \Var\!\left(\sum_k {x_k^\ast}^2\right) \approx 4\sigma^2\sum_k {x_{0,k}^\ast}^2=  4\sigma^2\mathrm{AMD_{e,0}}.
\end{equation}
It results that $\std (\AMDe)=2\sigma\sqrt{\mathrm{AMD_{e,0}}}$, \AMDi being constant in this model, we have
\begin{equation}
    \frac{\std(\AMD)}{\langle\AMD\rangle} \approx \frac{\sqrt{\Lambda}e_{\mathrm{floor}}}{\sqrt{2\AMD}}.
    \label{eq:AMD_trend}
\end{equation}
The latter trend is shown in Fig. \ref{fig:Global_No_symmetry}, using the median, over the planets and systems of each \texttt{Peas} setup, of $\Lambda$ and $e_{\mathrm{floor}}$.\\

We also observe, in Fig. \ref{fig:Global_No_symmetry}, a linear trend in the evolution of the median standard deviation of the \AMDi in the N-body dynamics and \AMDe with both secular and N-body integration methods. 
Given that we observe the same trend in both integrations, it must be a manifestation of the secular chaos, involving terms who violate the conservation of the quasi-integrals.
There is no simple diffusion model allowing us to precisely study these variations, however, we can search for a scaling law.
With the same reasoning as before we write
\begin{equation}
\mathbf{x}^\ast \approx\mathbf{x}^\ast_0+\mathbf{x}^{(4)}_{\mathrm{sec}}+\mathbf{x}^{(6)}_{\mathrm{sec}}, \quad \mathbf{y}^\ast\approx\mathbf{y}^\ast_0+\mathbf{y}^{(4)}_{\mathrm{sec}}+\mathbf{y}^{(6)}_{\mathrm{sec}}\label{eq:x_0_sec}
\end{equation}
where the components of $(\mathbf{x}_{\mathrm{sec}}, \mathbf{y}_{\mathrm{sec}})$ are correlated and the terms $\mathbf{x}^{(4)}_{\mathrm{sec}}, \,\mathbf{y}^{(4)}_{\mathrm{sec}}$ and $\mathbf{x}^{(6)}_{\mathrm{sec}}, \,\mathbf{y}^{(6)}_{\mathrm{sec}}$ denote the contribution to the mode evolution associated respectively with the order 4 and 6 terms of the secular Hamiltonian.
Since the order 4 terms in the secular Hamiltonian do not contribute to variations of \AMDe and \AMDi (see Fig. \ref{fig:AMDi_conserving_vs_non_conserving_freqs}), we can consider that the secular-chaos induced variations are dominated by order 6 secular resonances.
In the pendulum approximation, widths of the resonance islands are proportional\footnote{More precisely, the widths of the islands originating from order 6 resonances are proportional to the square root of monomials of degree 6 in eccentricity and inclination.} to $\AMD_0^{3/2}$.
The variance of $x^{(6)}_{\mathrm{sec},k}$ is directly proportional to the size of the resonant islands, thus $\Var(|x^{(6)}_{\mathrm{sec},k}|)\propto \AMD^3$.
We assume that such scaling will be present in all terms, neglecting the cancellation that could arise from covariances (as is the case for the total \AMD, conserved by secular dynamics at all orders, \citealp{Laskar_2017}), we have
\begin{equation}
    \std(\AMDe)\propto \sqrt{\mathrm{AMD_{e,0}}} \AMD_0^{3/2}.
\end{equation}
This reasoning can only be carried with orders of magnitudes since there exists strong correlation between the secular variations.
Finally, one can write
\begin{equation}
    \frac{\std(\AMDe)}{\langle\AMD\rangle} \propto \AMD.
\end{equation}
The same scaling holds for the standard deviation of \AMDi.
We note that not seeing variations in the modes with weaker scaling is a hint that the $4^\mathrm{th}$ order network is not sufficient to induce chaos for the considered systems configurations\footnote{The diffusion timescale is given by the lowest order resonances required for resonance overlap \citep{Petit_2020}. Because we do not see a scaling in $\AMD^{1/2}$, we can consider the order 4 is not sufficient to allow for large scale diffusion as the order 4 network may not be fully overlapped. Its contribution speeds up the diffusion of the modes but does not affect the quasi-integrals. From the recovered scaling, we suppose that the lowest order resonances for resonance overlap to be of order 6.}.

\subsection{\texttt{Far} set}
\begin{figure}%
\begin{center}
\includegraphics[width=1.0\linewidth]{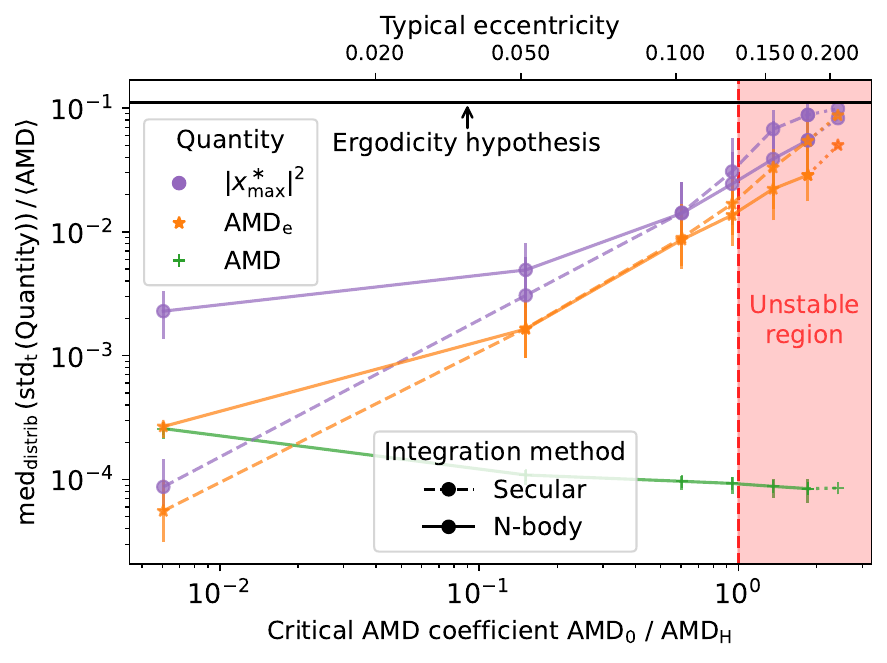}
\includegraphics[width=1.0\linewidth]{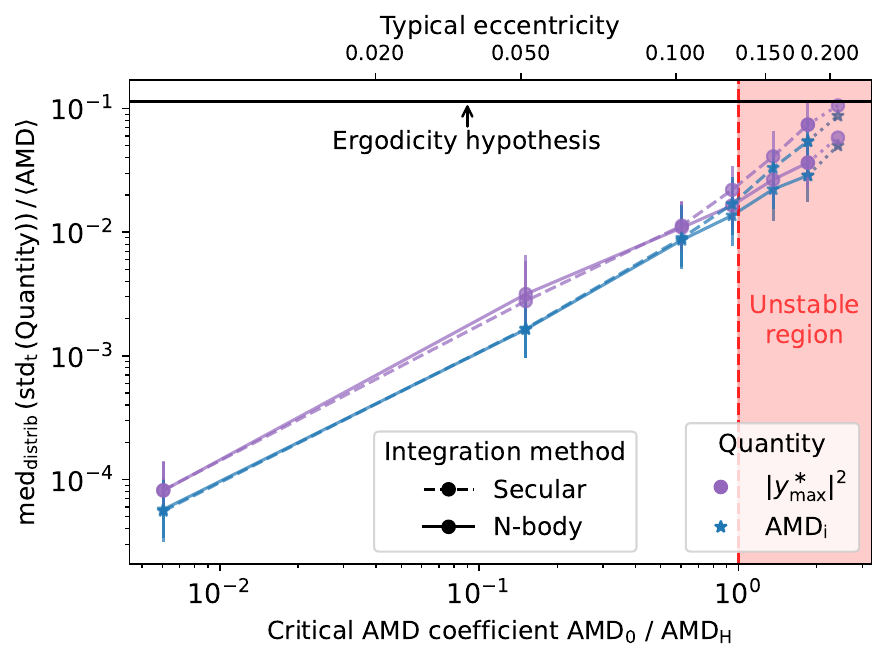}
\caption{Same as in Fig. \ref{fig:Global_No_symmetry}, for the \texttt{Far} set. 
\label{fig:Global_far}}
\end{center}
\end{figure}

We performed the same process for the 7 "\texttt{Far}" setups presented in Table \ref{tab:Simulation_params}, the resulting trends of standard deviations of \AMD, \AMDi, \AMDe and squared modulus of proper modes from the N-body and secular simulations are shown in Fig. \ref{fig:Global_far}.
The conservation of the AMD components is of the same order of magnitude as in the \texttt{Peas} setups, i.e. getting to the order of the percent at stability limit.
However the proper modes are much better conserved, in fact they are roughly as conserved as the \AMDe and \AMDi. This can easily be explained by the architecture of the \texttt{Far} systems.
Indeed, planets' period ratios are much larger and thus planet-planet interactions are not as strong. 
Frequencies are much more spread so it is harder for the resonance condition to be satisfied and thus less variations are expected to occur.
Again we notice that at high AMD we approach the ergodicity hypothesis limit.
Results are again fairly equivalent via both codes.
The difference between the distribution computed via the integrations of both codes start to be non-negligible for the \texttt{Far}$_6$ and \texttt{Far}$_7$ setups due to the non secular nature of these systems.
Again, at very low \AMD value, the large relative variations of \AMD is due to eccentricity kicks, and is thus due to \AMDe variations.
Planets being further away, the random eccentricity kicks at conjunction have a less pronounced effect.
Once again, the quasi-integral nature of both the inclination and eccentricity \AMD for systems within the \texttt{Far} setups is obvious.

We point out that we performed simulations of systems with a higher number of planets (see Sect. \ref{sec:High_Multiplicity}), smaller period ratios and different masses (consistent with super-Earth systems) and found again excellent conservation of the \AMDe and \AMDi (and of proper modes).

\subsection{Ergodicity versus quasi-integrals}
In the proper modes stability context, the normalized-std value of 0.11 computed under the ergodicity hypothesis appeared as a good approximation of the maximum diffusion that systems could reach.
Indeed, standard deviations approached this limit only for a priori unstable systems.
One could argue that going to higher critical AMD coefficients would allow to go beyond this limit, however no less than about 50\% of systems were lost in the \texttt{Peas}$_7$ setup ran via N-body simulations.
Thus, simulating the evolution of systems with higher \AMD and trying to see whether the ergodicity limit is significantly crossed would not be meaningful.
However the typical standard deviations of \AMDe and \AMDi were about one order of magnitude smaller than the one obtained under the ergodicity hypothesis, even for a priori unstable systems.
This difference of behavior between the proper modes and the \AMDe and \AMDi in the \texttt{Peas} systems is once again an argument in favor of the existence of quasi-integrals and thus of a more restrictive dynamical model where, for stable systems, the evolution is restrained by the quasi-integrals.

\section{Discussions}\label{sec:discussions}
We have shown, in the previous section, the existence of the \AMDe and \AMDi quasi-integrals of motions for super-Earth systems of planets.
We discuss here the consequences of these quasi-integrals on the longterm stability of planetary systems and the implications on the dynamical history of systems.
We also comment on the observed secular diffusion and the relevance of secular chaotic dynamics.

\subsection{\texorpdfstring{The $e/i$ ratio as a dynamical tracer of system history}{The e/i ratio as a dynamical tracer of system history}}
Recent work has shown that the eccentricities and inclinations of exoplanets are typically of a few degrees and a few percents, respectively (see Sect. \ref{sec:Introduction}).
As seen through our simulations, in this range, the \AMD-normalized standard deviations of the \AMDe and \AMDi is typically below the percent.
More generally, the quasi-integrals imply that there exists a double reservoir of \AMD. The existence of two separate reservoirs in eccentricities and inclinations suggests that the total eccentricity and inclination excitations of systems is conserved.
This is significant, since one can, via the eccentricities and inclinations observables, link the current state of a system to the system's last reorganization phase.
The latter could be the disk dissipation, a violent collision phase or an important tidal damping.
All these mechanisms leave a different imprint onto the quasi-integrals and in particular onto the \AMDe/\AMDi ratio.
A system primarily shaped in a protoplanetary disk which broke from a resonant configuration will tend to have a very large $e/i$ ratio \citep{Izidoro_2021}.
The dynamical reorganization gives a more balanced $e/i$ ratio spanning from 1 to 2 \citep{Kokubo_2025}, where the exact value depends on how extended the system is.
On the other hand if tides were the last reshaping mechanism at play, eccentricities will be significantly damped leaving inclinations unaffected.

While the evolution of eccentricities and inclinations driven by gravitational interactions has been extensively studied in planetesimal discs \citep[e.g. in][]{Ida_1993,Ohtsuki_2002}, the corresponding distributions  in planetary systems have received less attention, with \citet{He_2020} examining them in the context of AMD stability. Our results build on this work by highlighting the potential of the \AMDe/\AMDi ratio as a probe of planetary systems' dynamical histories, offering a way to constrain the processes that last shaped their architectures.

\subsection{Implications on AMD-stability}
    
\begin{figure}%
\begin{center}
\includegraphics[width=1.0\linewidth]{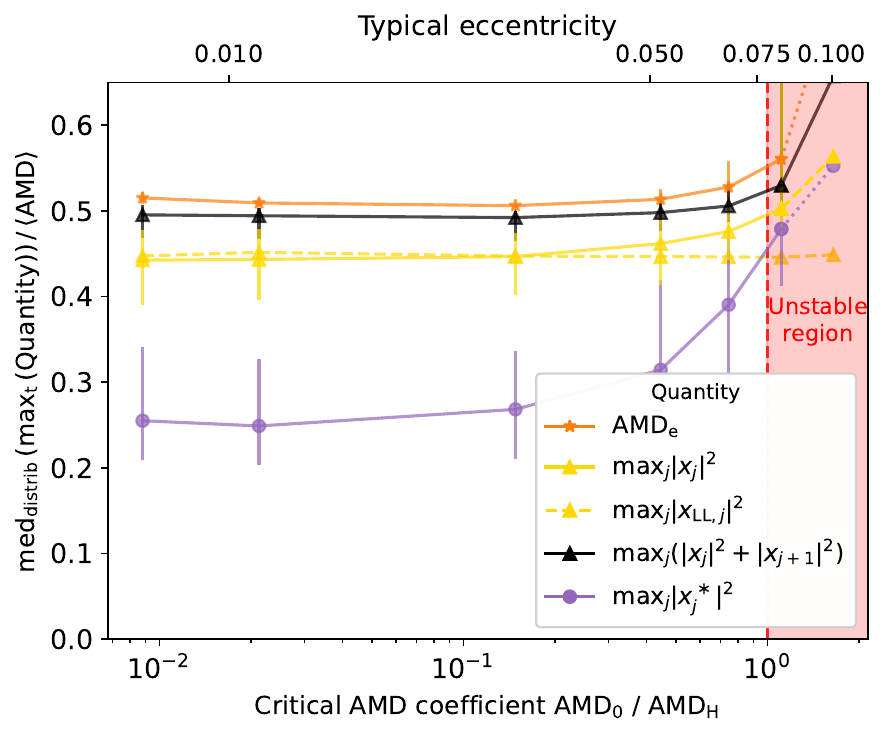}
\caption{Median, over the distribution, of the temporal maxima of: \AMDe, squared modulus of the eccentricity proper modes, $x$ Poincar\'e  variable, $x$ Poincar\'e  variable assuming LL dynamics (i.e. reconstructed from Eq. \eqref{eq:Poincare_reconstructed}) and \AMDe in a pair of planet for the N-body \texttt{Peas} simulations.
For each quantity (except \AMDe) we consider the maximum value, over the planets or planet pairs.%
}
\label{fig:Global_No_symmetry_max}
\end{center}
\end{figure}

Another consequence of the quasi-integrals nature of \AMDe and \AMDi is on the a priori stability of systems.
In the work presented in \cite{Laskar_2017}, it was assumed that AMD-stability for a pair of planets was satisfied when the orbits of the two planets could not intersect under the assumption the \AMD was fully contained within that pair.
This uses the fact that the amount of \AMD contained in the inclination degree of freedom can be freely converted into eccentric \AMD, i.e. exchanges are fully allowed. The stability was then considered ensured for systems with $\AMD<\mathrm{AMD_{crit, intersect}}$.
However, our study suggests that for a priori longterm-stable systems, there are very little exchanges between \AMDe and \AMDi. 
Therefore the criterion for collisions in the AMD framework should consider stability when the orbits cannot intersect under the assumption that \AMDe is fully contained within the pair.
The criterion should thus read $\AMDe<\mathrm{AMD_{crit, collision}}$.
In the particular case where $\AMDi=0$ then the criterion of \cite{Laskar_2017} is unchanged, otherwise it is too restrictive and consequently systems previously identified as \AMD-unstable might be in fact \AMD-stable.
This is illustrated by Fig. \ref{fig:Global_No_symmetry_max}, which shows that the maximum amount, over the 100 Myr evolution, of \AMDe within a pair ($j,j+1$) of planets, i.e. $\mathrm{max}_j(|x_j|^2 + |x_{j+1}|^2)$, reaches 50\% of the total \AMD.
In other words a pair of planets can very well carry the entire \AMDe available in the system, despite the relatively good conservation of the modes.
Planets can indeed exchange \AMDe between them, even to the point where a planet pair possesses the entire \AMDe of the system, however the \AMDi is never significantly converted into \AMDe.

However, so far, we considered specific systems assuming that the AMD was equipartitioned between the eccentricity and inclination degrees of freedom, with roughly equipartition of the AMD within planets, "only" 4 planets of same mass and with a small spread of adjacent planets' period ratios.
Even though this represents a typical compact super-Earth system, it is only covering a very small portion of the exoplanetary systems' phase space.
We expect the conservation to be different depending on the architecture of systems.
In Sect. \ref{sec:diffusion} and Sect. \ref{sec:High_Multiplicity}, we provide a preliminary exploration of these dependencies by respectively considering systems where AMD is not equipartitioned between planets and considering systems with higher multiplicity.
In the particular case of "hierarchical" systems, i.e. with inner and outer subsystems there has not been any global study, it is not clear how the \AMD and its components transfer between the inner and outer parts.
The only study conducted was for the Solar System, where \citet{Mogavero_2023} showed that the inner system's \AMDi is quasi-conserved and that the \AMDe and \AMD are not. 

\subsection{Longterm secular diffusion}\label{sec:diffusion}

\begin{figure}%
\begin{center}
\includegraphics[width=1.0\linewidth]{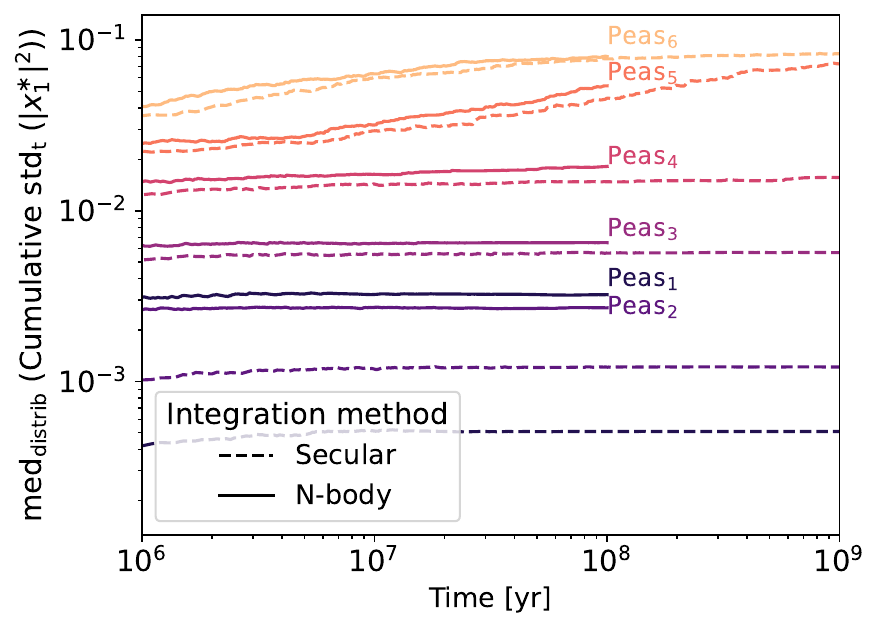}
\caption{Median, over the distribution, of the cumulative temporal standard deviation of the first eccentricity mode for the \texttt{Peas}$_1$ to \texttt{Peas}$_6$ systems ran via both N-body and secular evolution codes.}\label{fig:Cumulative_std_Peas}
\end{center}
\end{figure}

We show in Fig. \ref{fig:Cumulative_std_Peas} the typical cumulative temporal standard deviation of an eccentricity proper mode in the \texttt{Peas}$_1$ to \texttt{Peas}$_6$ setups, obtained via both numerical codes.
In the \texttt{Peas}$_1$ to \texttt{Peas}$_3$ setups, we do not see any evolution in the cumulative standard deviation of the mode and the values are very small.
The diffusion is very limited, it saturates very rapidly. 
Despite the chaotic nature of the dynamics, as evidenced by a nonzero Lyapunov exponent, the chaotic region of the phase space explored by these systems is so small that they behave practically as if they were quasi-periodic.
The \texttt{Peas}$_4$ setup appears to be the transitional setup between confined and large scale secular chaos, as we notice a very slight increase in the cumulative std.
Indeed, we do not reach a plateau in std, it appears the diffusion is still slightly active at the end of the integration time, even after 1 Gyr.
However the increase is very small.
We found that the typical Lyapunov timescale of the \texttt{Peas}$_4$ is about 1.4 Myr.
For systems sufficiently below instability limit, it appears that integrating further in time does not bring out any significant feature in the dynamics.
For the \texttt{Peas}$_5$ and \texttt{Peas}$_6$ systems we clearly have an increase in the cumulative std of the mode and the diffusion has not reached a steady state at the end of the integration time.
When increasing the \AMD, we transition from the almost quasi-periodic regime with confined secular chaos to the large scale globalized secular chaos.
Yet, the conservation of \AMDe and \AMDi is still of the order of the percent for most \texttt{Peas}$_5$ and \texttt{Peas}$_6$ systems.

Our secular integrations simulations are performed over timescales of at least an order of magnitude larger than the  N-body integrations.
In the case of the \texttt{Peas} series, secular integrations went up to 1 Gyr, i.e. more than 30 billion orbits of the inner planet.
We could have expected the results from the two integration methods to be different, as dynamical features could very well emerge only on very long timescales.
However, we notice that both the secular and N-body codes give very similar results in the regime where secular dynamics dominates, i.e. everywhere except at both very low \AMD (for the eccentricities only)
and very high \AMD.
One could thus perform very longterm simulations of planetary systems and then study the \AMD components and proper modes evolutions using secular integrations with minimal loss of information.

\begin{figure}%
\begin{center}
\includegraphics[width=1.0\linewidth]{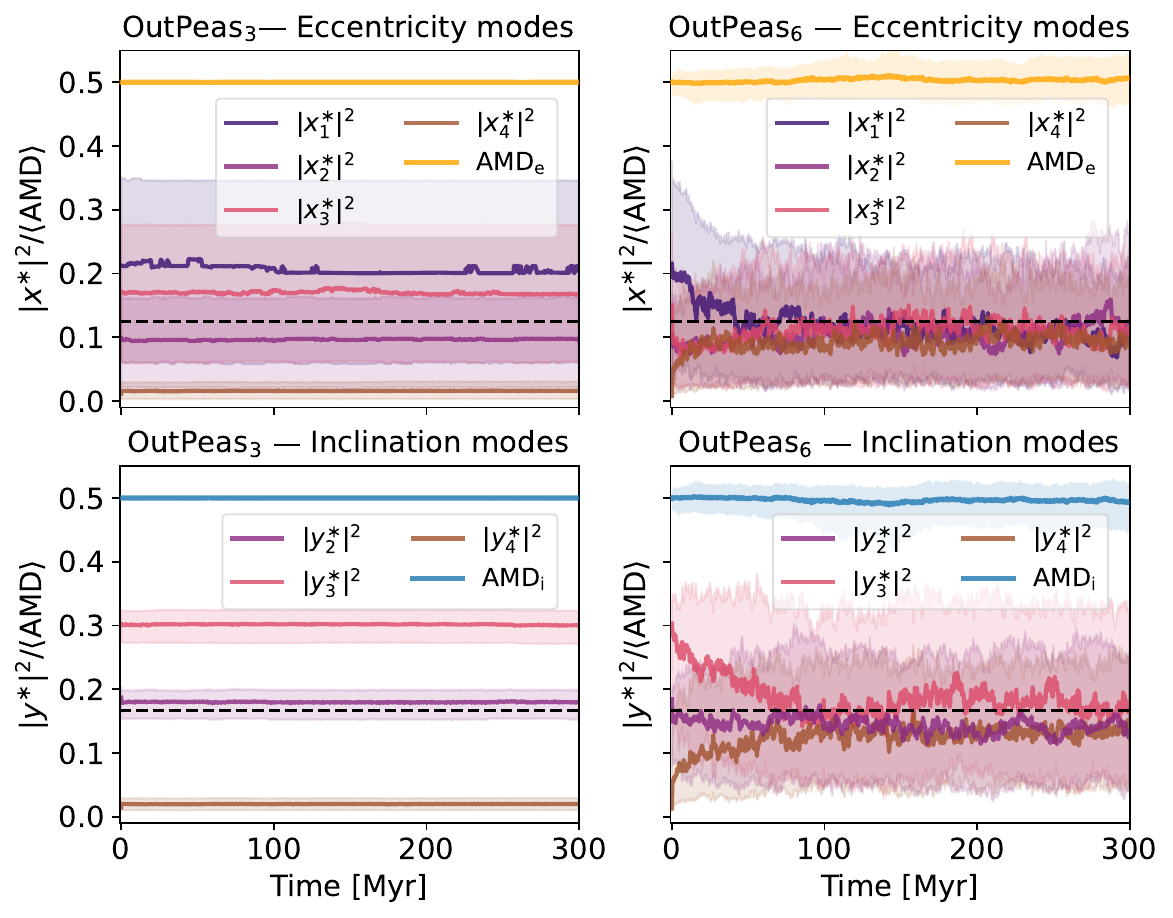}
\caption{Median over the 80 systems of the evolution of the AMD-normalized squared Laplace-Lagrange modes, the \AMDi and \AMDe, over 300 Myr for systems belonging to the \texttt{OutPeas}$_3$ and \texttt{OutPeas}$_6$ setups (respectively left and right subplots).
The time series are low-pass filtered using a cutoff frequency of 0.5 Myr via a 4$^{\text{th}}$ order Butterworth filter. The mode $y_1^\ast$ associated to the null frequency $s_1$ is not shown as its amplitude is null at all times. The black horizontal lines are at the theoretical equipartition levels of the modes. The filled areas represent the $15^{\mathrm{th}}$ and $85^{\mathrm{th}}$ percentiles of the data, which for \texttt{OutPeas}$_3$ correspond to the spread of initial conditions.}\label{fig:OutPeas}
\end{center}
\end{figure}

Our \texttt{Peas} systems were initialized at thermodynamical equilibrium, in the sense that the \AMD was uniformly distributed between the four planets. 
To test the soundness of this assumption, we performed simulations, called \texttt{OutPeas}$_3$ and \texttt{OutPeas}$_6$ (see Table \ref{tab:Simulation_params}), where the total \AMD of the systems is the same as in \texttt{Peas}$_{3}$ and \texttt{Peas}$_{6}$ (respectively) but with the total \AMD only distributed between the two outermost planets.
In this setup, the proper modes do not initially have similar amplitudes.
More precisely the distribution of \AMD between the planets is such that one mode initially dominates, one is at the equipartition level, and the remaining modes of eccentricity (resp. mode of inclination) are below equipartition level (resp. is close to 0). 
Since from one system to another the repartition is roughly equivalent, we can apply a median over systems to study the ensemble behavior.
We show in Fig. \ref{fig:OutPeas} the evolution of the median, over systems, of the eccentricity and inclination proper modes as well as \AMDe and \AMDi (i.e. their respective sums) from these setups over 300 Myr. 

The \texttt{OutPeas}$_3$ systems do not show any variation of the modes amplitude over 300 Myr. We interpret this phenomenon from the almost regular LL behavior of their dynamics. As these systems are only mildly chaotic and  no \AMD transfer occurs between mode, the mode amplitudes are conserved, as in LL dynamics.
We conclude that, as secular chaos is too weak, the longterm evolution of LL-like systems preserves the initial distribution.
On the other hand,  for the \texttt{OutPeas}$_6$ setup, the medians over the 80 systems, of the mode distributions reach equipartition after a few tens of million years.
We interpret these exchanges of modes as a consequence of the longterm diffusion due to secular chaos driving the systems' evolution and allowing complete redistribution of modes, towards equipartition.
A given choice of initial mode distribution therefore does not seem to have a significant impact, as secular chaos will simply redistribute modes if they are not initially close to equipartition.

Despite the redistribution of modes, the \AMDe and \AMDi of the systems are well conserved.
The medians, over the 80 simulated systems, of the \AMD-normalized temporal std of \AMDe and \AMDi are $\approx 0.001$ for \texttt{OutPeas}$_3$ and $\approx0.015$ for \texttt{OutPeas}$_6$. 
These values are consistent with the stds found with the \texttt{Peas}$_3$ and \texttt{Peas}$_6$ systems with roughly uniform distributions of \AMD. 
Therefore we conclude that our results on the existence of quasi-integrals are independent of the considered system's initialization.

\subsection{Relevance of secular chaos for Peas systems}

We show in Fig. \ref{fig:Global_No_symmetry_max} another measure of chaos by considering the distribution median of the maximum (over time) values reached by $|x^{\ast}|^2$ and $|x|^2$ and maximum values of \AMDe in the entire system, over the 100 Myr N-body integrations.
Figure \ref{fig:Global_No_symmetry_max} shows that, as expected, in the case of \AMDe, the median value is roughly equal to the initial equipartition level, i.e. we have $\AMDe/\AMD\approx0.5$, for systems below the instability threshold. 
In other words, the exchanges between \AMDe and \AMDi were never sufficient to significantly break the initial equipartition.
In the case of the proper modes, (in purple in Fig. \ref{fig:Global_No_symmetry_max}), the median is only $\sim 25\%$, i.e. at equipartition level.
This is consistent with the time series shown in Fig. \ref{fig:OutPeas}, where mode distributions are conserved at low \AMD.
Since it starts from a relatively low value, it has a lot of flexibility to increase and as secular chaos becomes more prominent, it reaches the maximum allowed by the almost conservation of \AMDe.
For the squared $x$ Poincar\'e  variables, the median over the distribution is roughly 45\% and increases much slower.
This quantity is equivalent to the \AMD contained in one planet.
Therefore it seems that the system can, at some point during the evolution, reach a state where $90\%$ of the \AMDe is concentrated within one planet. This is particularly meaningful because the dynamics are here close to the Laplace-Lagrange regime.
To assess this, we compare the $x$ Poincaré variables obtained from the N-body evolution with those reconstructed under the Laplace-Lagrange approximation ($x_\mathrm{LL}$).
The latter variable is obtained by expressing the initial conditions as a linear combination of the LL eccentricity modes, whose amplitudes are conserved in the LL framework, and evolving each mode at its corresponding secular frequency (see Eq. \eqref{eq:Poincare_reconstructed}).
The resulting ratio of the AMD-normalized std of $|x_\mathrm{LL}|^2$ is close to that obtained from N-body integrations, suggesting that significant exchanges of \AMD between planets can occur even in systems whose dynamics remains close to the Laplace-Lagrange regime, allowing up to 90\% of the \AMDe to be concentrated in a single planet.

Moreover we find that the median, over the distribution, of the temporal standard deviation of mean$_j(|x_j|^2/\langle \AMD \rangle)$, i.e. the typical AMD-normalized standard deviation of the amount of \AMDe in a single planet, ranges from 0.08 (low critical \AMD coefficient) to 0.10 (high critical AMD coefficient).
These values are extremely close to the one computed under the ergodicity hypothesis ($\sim 0.11$). 
In other words each planetary eccentricity explores the entire space that is available at the \AMDe value. 
Even though planets exchange a large amount of \AMDe, it appears the latter is very well conserved.
Despite the conservation diminishing at higher \AMD values, modes are globally well conserved.
Secular chaos does not seem to bring out any significant dynamical behavior for the eccentricities and inclinations for our \texttt{Peas} systems, where modes are initially close to equipartition.
The projection of the quasi periodic motion onto a single eccentricity variable already covers the entire available space.
This is however not a general feature since particular mode initialization can differ from the ergodic distribution \citep{Laskar_2008, Mogavero_2017}.

\subsection{Secular chaos and quasi-integrals in high multiplicity systems}\label{sec:High_Multiplicity}

Our study has so far been carried out for 4-planet systems only. We explore here the importance of secular chaos and the conservation levels of \AMDe and \AMDi for high multiplicity peas systems. We thus performed 80 simulations of 8-planet systems, (\texttt{8Peas} setup in Table \ref{tab:Simulation_params}).
The \AMD is an extensive quantity, if we increase the number of planets while keeping the AMD constant, we dilute the \AMD across more planets.
To keep a comparable level of dynamical excitation as our four planet systems,
we chose to consider the same typical eccentricity and inclination as in the \texttt{Peas}$_5$ setup.
For these simulated 8-planet systems, the ratio $\AMD/\mathrm{median}(\AMDH)$ is roughly 2.26 compared to 0.75 for the  \texttt{Peas}$_5$ systems.
After 100 Myr, only 19 of the 80 systems integrated have become unstable.
Due to the high multiplicity, it is in fact harder for a given planet to concentrate enough AMD to become Hill unstable.
The ratio of the orbital eccentricity to the orbit crossing eccentricity, $\frac{e}{e_{\mathrm{cross}}}=1-\alpha$, decreases for pairs of outer planets, suggesting that the effect of chaos are less pronounced for these pairs.
However, since the \AMD is distributed between eight planets, the secular interactions lead to chaos via longterm diffusion and thus the longer is the integration the more a (most likely inner) planet can eventually become unstable after concentrating enough \AMD to reach orbit crossing.
This chaotic diffusion produces late instabilities as after 600 Myr only 37 systems are still stable.
Therefore for these eight planet systems, it appears that secular chaos does matter.

\begin{figure}%
\begin{center}
\includegraphics[width=1.0\linewidth]{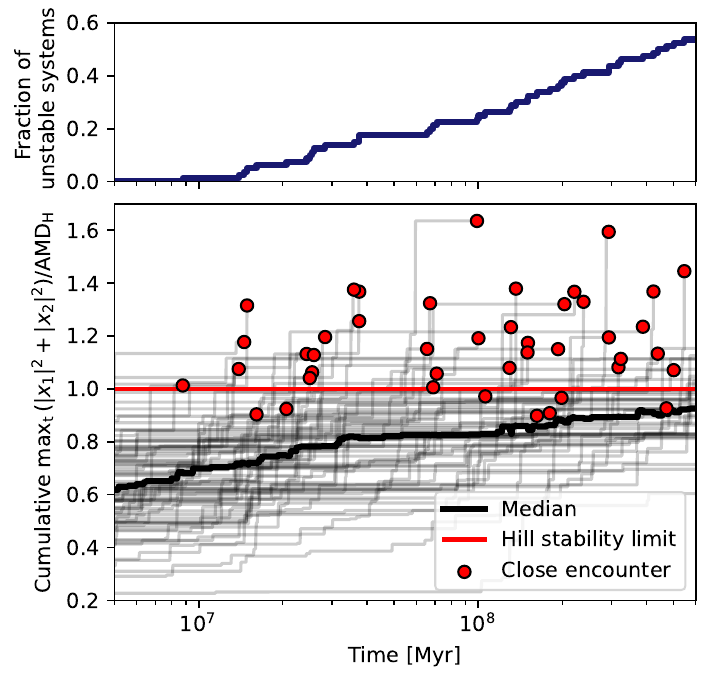}
\caption{Cumulative temporal maximal amount of \AMDe in the inner pair of the \texttt{8Peas} systems, normalized by the Hill stability criterion for these systems. The instability times of unstable systems are represented by red disks. Unstable systems are kept for the computation of the median to reflect the ensemble dynamics.}\label{fig:Cumulative_inner_pair_AMD}
\end{center}
\end{figure}

We show in Fig. \ref{fig:Cumulative_inner_pair_AMD} the temporal maximal amount of \AMDe that is contained in the inner pair of planets for each system of the \texttt{8Peas} setup.
We clearly see that the inner pair reach higher and higher \AMDe with time, due to the secular chaotic diffusion of \AMDe.
As the pair's \AMDe increases and the orbits eventually cross, planets have close encounters and systems become unstable.
We note that both \AMDe and \AMDi are well conserved in the integrated systems, with a median value of temporal std at percent level. 
Comparing these 8-planet systems with our \texttt{Peas} systems suggests that the number of degrees of freedom influences the stochastic evolution of the \AMD redistribution among the planets, as it increases, the evolution becomes anisotropic.
This effect was already observed, for instance in \cite{Volk_2020}, which focuses on the five-planet Kepler-102 system and finds that, again, inward \AMD transfer due to secular chaos can lead to planet-planet collisions by exciting the inner planet's eccentricity.
We intend on studying this phenomenon more thoroughly in a future paper.

\subsection{Secular chaos beyond peas in a pod}
The very good conservation of \AMDe and \AMDi only acts as a barrier in terms of stability.
Based on our \texttt{Peas}, \texttt{OutPeas} and \texttt{Far} setups, the chaotic nature of a 4-planet system appears to have little impact on its evolution, regardless whether the \AMD is initially well distributed or not.
Secular chaos does in fact not seem relevant for typical "peas in a pod" systems \citep{Weiss_2018}.
However, for non peas in a pod systems, with large disparities in the initial distribution of \AMD, where a flux of \AMD (due to secular chaos) from one part of the system to another occurs, chaos can be dynamically important.
This is the case of the Solar System where the inner system instability is caused by an AMD transfer from the outer one after the activation of the $g_5-g_1$ resonance \citep{Laskar_2008,Hoang_2022}.
Beyond the Solar System, \citet{Wu_2011} looks at cases of 3-planet (with Jupiter-like masses) systems where the \AMD of outer planets is larger than the angular momentum of the inner planet.
This choice of initial condition results in an important inward transfer of AMD which tends to destabilize the system, as a change in the \AMD of an inner lower mass planet has a great effect on its eccentricity (and inclinations).
Similarly, \cite{Petrovich_2019} considers super-Earth systems extending to a few AUs with $N>3$ planets and where the innermost planet is much less massive than the outer ones.
Again, the outer planets carry much more \AMD and secular chaos drives the inner planet to higher \AMD values, leading to inward tidal migration and consequently making them ultra-short-period planets via later circularization.

\section{Conclusions}
The recent work on the longterm dynamics of the Solar System, conducted by \citet{Mogavero_2023}, has shown that the evolution of the inclination Angular Momentum Deficit (\AMDi) of the inner planets, describing the planets' mean mutual inclination, is quasi-conserved.
In this paper, we provided a first numerical exploration of the \AMDi and \AMDe conservations beyond the Solar System.
We studied different configurations of systems, from close-in packed super-Earth systems to further out and spread out systems, and found that conservation holds in both cases.
We generalized the results from \citet{Mogavero_2023} showing that the \AMD-components are well conserved over secular timescales.
We give an explanation for the generality of this conservation, proposing that a significant part of the web of secular resonances leaves the \AMDe and \AMDi unaffected. One can thus consider both \AMDe and \AMDi as quasi-integrals of motion.
We also showed that the proper modes become only fully affected by chaotic diffusion near the crossing condition. 
At lower AMD values, the modes have standard deviations roughly an order of magnitude larger than the \AMDe and \AMDi. 
This is a striking result considering that the \AMDe and \AMDi are, respectively, the sums of the squared amplitudes of the eccentricity and inclination modes.
Unstable systems however do show full scale chaos for the secular degree of freedom.
The existence of the quasi-integrals reduces the amount of degrees of freedom for systems, and thus constrain their evolutions and allow us to probe the dynamical history of systems.
However, in the context of the PLATO and Gaia missions, it seems relevant to explore this conservation study to even more various systems.
We performed a preliminary exploration of the conservation in 4-planet peas in a pod systems where the \AMD is not initially at equipartition, as well as high multiplicity peas in a pod systems, and found that again quasi-conservation holds.
For high multiplicity systems secular chaos does have an impact on the secular dynamics.
It would be of interest to perform other simulations with a variety of planetary masses and with different architectures, including hierarchical systems, extending the work of \citet{Pu_2018} and \citet{Pu_2021}.
We intend, in the future, to investigate the conservation for such systems and, if possible, to trace back their dynamical history.

\begin{acknowledgements}
We thank Daniel Tamayo, reviewer of this manuscript, for his constructive remarks that significantly improved this paper.
The authors are grateful to the OPAL infrastructure from Observatoire de la Côte d'Azur (CRIMSON) for providing resources and support.
This work was supported by the French government, through the $\mathrm{UCA^{J.E.D.I.}}$ Investments in the Future project managed by the National Research Agency (ANR) with the reference number ANR-15-IDEX-01.
Simulations in this paper made use of the REBOUND N-body code \citep{Rein_2012}. The simulations were integrated using the WHFast and WHFast512 symplectic Wisdom-Holman integrators \citep{Rein_2015,Dagli_2026,Wisdom_1991}. The Simulationarchive format was used to store fully reproducible simulation data \citep{Rein_2017}.
This research was also made possible by the open-source projects Jupyter \citep{Kluyver_2016}, numpy \citep{VanDerWalt_2011}, matplotlib \citep{Hunter_2007}, xarray \citep{Hoyer_2017} and celmech \citep{Hadden_2022}.
\end{acknowledgements}

\bibliographystyle{aa}
\bibliography{e_i_decoupled.bib}

\begin{appendix}
\nolinenumbers

\section{Simulation alignment procedure}\label{appendix:initialization_systems}
As explained in Sect. \ref{sec:Secular_dynamics}, it is practical to have the angular momentum $\mathbf{L}$ aligned with the $z$-axis.
There are different ways to make sure this condition is satisfied, in our case we chose a model using springs.
We consider $N-1$ mass points connected by springs of sizes $L_j$ and hooked at the origin $(0,0,0)$ and at a point $(0,0,L_z)$. 
Denoting the momentum-coordinate conjugate pair ($\mathbf{p}, \mathbf{r}$) the Hamiltonian of this system is
\begin{equation}
	\mathcal{H}=\sum_{j=1}^{N-1}\frac{\mathbf{p}_j^2}{2m}+\sum_{j=1}^N\frac{k}{2}\left(||\mathbf{r}_j-\mathbf{r}_{j-1}||-L_j\right)^2,
	\label{eq:H_springs}
\end{equation}
where $k$ is the springs stiffness constant.

The time derivative of the momenta are given by Hamilton's equations and reads, including a dissipation force proportional to velocity $v$,
\begin{equation}
\begin{aligned}
\frac{d \mathbf{p}_j}{dt} &= -\lambda \frac{\mathbf{p}_j}{m} - k(\Delta r_{j,j-1}-L_j)\frac{\mathbf{r}_{j}-\mathbf{r}_{j-1}}{\Delta r_{j,j-1}} \\
&\quad + k(\Delta r_{j+1,j}-L_{j+1})\frac{\mathbf{r}_{j+1}-\mathbf{r}_{j}}{\Delta r_{j+1,j}},
\end{aligned}
\end{equation}
where $\lambda$ is the friction force parameter, every vector is 3D and $\Delta r_{j,l} = \|\mathbf{r}_{j}-\mathbf{r}_{l}\|$.

We increase the spring constant as $k(t)=k_0\max(1,t/t_{relax})$ to stiffen the system. The idea is that the solutions of this problem give $N$ vectors with the right norm of the sum.

Moreover the results shown in this paper have been obtained setting $\lambda$ to $1$, $k_0$ to $0.1$, the relaxation time to $10$ and finally the masses of the bodies attached to springs to $1$ for simplicity.

We randomly draw the components of a vector $\mathbf{A} \in \mathbb{R}^{2N}$ from a standard normal distribution. The vector is normalized and then scaled such that its squared norm corresponds to the target \AMDe.
This results in a vector on the surface of the hypersphere $\mathbb{S}^{2N-1}$ of radius $\sqrt{\AMDe}$ embedded in $\mathbb{R}^{2N}$.

The scaled vector is then converted into $N$ complex Poincar\'e  variable $x_j$. %
The planetary eccentricities $e_j$ are then very simply computed as
\begin{equation}
    e_j=\left|x_j\sqrt{\frac{2}{\Lambda_j}\left(1-\frac{x_j\bar{x}_j}{2\Lambda_j}\right)}\right|.
\end{equation}
From these eccentricities, the actual angular momenta are computed as $L_j = \Lambda_j \sqrt{1 - e_j^2}$, ensuring that the total eccentricity-driven AMD matches the target value.
These values define the equilibrium lengths $L_j$ of the springs in the initial system.
In order for the systems to have the desired \AMD, we must incorporate the target \AMD associated with inclination, $\AMD_{\mathrm{i}}$. We thus impose that the total $z$-component of the angular momentum, $L_z$, is reduced by this amount.
Explicitly, we set 
\begin{equation}
    L_z = 1- \frac{\AMD_{\mathrm{i}}}{\sum_j L_j}.
\end{equation}
This ensures that the component of angular momentum perpendicular to the reference plane reflects the desired level of mutual inclination in the system.
The intermediate nodes of springs (i.e. the mass points) have their initial transverse coordinates drawn from a Gaussian distribution with characteristic width chosen to be representative of the expected transverse extent of the chain for the prescribed value of \AMDi.
The springs are initialized with zero velocities.
As mentioned earlier, the spring system is constructed so that the endpoint of the chain lies at $(0,0,L_z)$.
The $z$ coordinate of intermediate nodes are set according to the cumulative sum of the planets' angular momentum fractions, ensuring that the \AMDi is distributed proportionally along the chain.
This forms the starting point for the dissipative evolution that leads the system to equilibrium.

After initializing the system, we solve the equations of motion derived from the Hamiltonian in Eq. \eqref{eq:H_springs}, including a damping force proportional to velocity.
The time evolution of the spring-mass system gradually relaxes toward a configuration in which the lengths of the spring segments match the prescribed angular momentum norms.
To facilitate this, we gradually increase the spring stiffness $k(t)$ as explained earlier.

Once the system reaches equilibrium, each segment of the spring chain corresponds to a planet's angular momentum vector.
From these vectors, we can directly compute the inclinations and longitudes of ascending nodes.
More precisely, let $\mathbf{L}_k = (L_{x,k}, L_{y,k}, L_{z,k})$ be the angular momentum vector of planet $k$ (and where ${x,y}$ refer to cartesian coordinates and not Poincar\'e  variables), recovered as the difference between two consecutive spring node positions after relaxation.
The planetary inclination is then given by
\begin{equation}
i_k = \arcsin \frac{\sqrt{L_{x,k}^2 + L_{y,k}^2}}{L_k} .
\end{equation}
The inclinations arise naturally from the spatial orientation of the spring vectors, that is, from how much each angular momentum vector tilts away from the $z$-axis.
This approach guarantees that the resulting system conserves total angular momentum and achieves the desired \AMD (both in eccentricity and inclination) in a statistically consistent way.
We can then easily compute $y$, the second complex Poincar\'e  variable. We therefore easily recover a system of $N$ planets whose initial orbital elements are set via the $x$ and $y$ Poincar\'e  variables obtained via the string model. 

\begin{figure}%
\begin{center}
\includegraphics[width=1.0\linewidth]{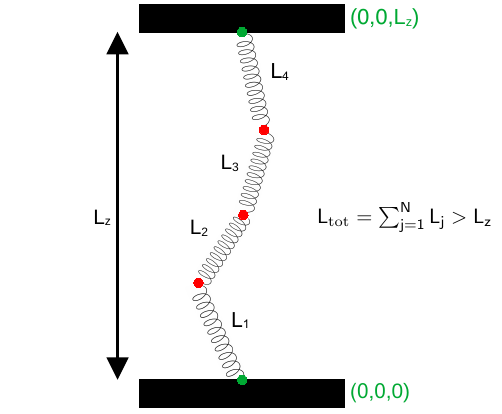}
\caption{Springs model (2 dimensional) representation after equilibrium is reached. The green disks represent the hooks at origin and at $(0,0,L_z)$ whereas the red disks are the mass points.}
\label{fig:Springs_sketch}
\end{center}
\end{figure}

\begin{figure}%
\begin{center}
\includegraphics[width=1.0\linewidth]{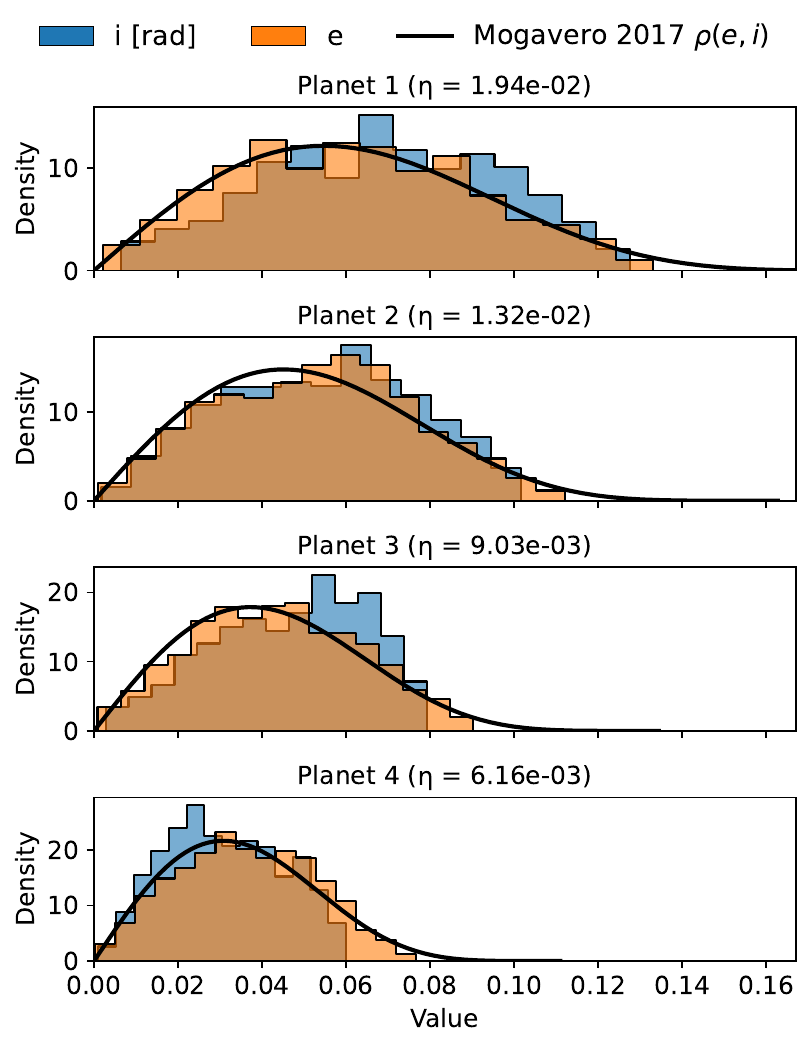}
\caption{Histograms of the initial eccentricity and inclination distributions obtained for 1000 initial conditions of 4 planet systems with \AMDe and \AMDi of the \texttt{Far}$_2$ setup.
The analytical PDF of eccentricity and inclination (given by Eqs. \eqref{eq:pdf_e} and \eqref{eq:pdf_i} respectively) is shown as a solid black curve.}
\label{fig:compar_strings_Mogavero_sampling}
\end{center}
\end{figure}

We show in Fig. \ref{fig:Springs_sketch} a sketch of one typical springs configuration obtained after equilibrium is reached and in Fig. \ref{fig:compar_strings_Mogavero_sampling} the eccentricity and inclination distributions obtained with the springs method for a large number of initial conditions. The distributions get narrower as we move to the outermost planet (planet 4).
This is consistent with the fact that the aim is to have planets with roughly the same contribution to the total AMD. Since $\Lambda_k$ grows as $\sqrt{a_k}$, one needs to have increasingly smaller inclinations and eccentricities of planets as we move from the innermost planet to the outer planet.
We compare the $e$ and $i$ distributions obtained via our springs method with the PDF used in \cite{Mogavero_2017} under the hypothesis of equiprobability in the phase space, which reads
\begin{equation}
    \rho(e_k)=\frac{2N-1}{\eta_k}e_k\left(1-\frac{e_k^2}{2\eta_k}\right)^{2N-2}, \quad e_k\leq \sqrt{2\eta_k},
    \label{eq:pdf_e}
\end{equation}
\begin{equation}
    \rho(i_k)=\frac{2N-1}{\eta_k}i_k\left(1-\frac{i_k^2}{2\eta_k}\right)^{2N-2}, \quad i_k\leq \sqrt{2\eta_k},
    \label{eq:pdf_i}
\end{equation}
and with $\eta_k=\AMD/\Lambda_k$.
Figure \ref{fig:compar_strings_Mogavero_sampling} shows that the distributions of eccentricities and inclinations obtained via the springs model are consistent with the analytical PDF.

\section{Computation of the modes, \texorpdfstring{\AMDe and \AMDi}{AMDe and AMDi} variability under the ergodic hypothesis}\label{appendix:ergodicity}

We present here the computation of the modes and \AMD components variability in the ergodic hypothesis, used in Sect. \ref{sec:Results} as an upper limit for the conservation.%

We randomly draw each components of a vector $\mathbf{B} \in \mathbb{R}^{4N}$ via a standard normal distribution, i.e. $B_i\sim\mathcal{N}(0,1)$. $\mathbf{B}$ is normalized and is thereby on $\mathbb{S}^{4N-1}$ (i.e. $\mathbb{S}^{15}$ in our case) and then scaled such that its squared norm corresponds to the total \AMD target, let $\mathbf{B'}$ be this new vector.
The x Poincar\'e  coordinates of the $N$ planets are then defined as $\mathbf{x}=\mathbf{B'}_{1,...,N}+\iota \mathbf{B'}_{N,...,2N}$, i.e. using the first $2N$ components of $\mathbf{B'}$.
The eccentricity \AMD is defined as the sum of the squared modulus of the $x$ Poincar\'e  coordinate and consequently the ratio $\AMDe/\AMD$ follows a beta distribution, with both parameters equal to $N$, i.e. $\AMDe/\AMD\sim$Beta$(N,N)$.
The remaining contribution to the total AMD is then equal to \AMDi and is used as input in the springs model (presented in Appendix. \ref{appendix:initialization_systems}).
In order to estimate the variability of proper modes, \AMDe and \AMDi in the ergodicity hypothesis, we draw 1 million initial conditions. 
Using the properties of ergodic dynamics where a temporal and space average are equivalent, we reshape the 1 million $\mathbf{x}$ and $\mathbf{y}$ such that we have an array equivalent to the evolution of 250 systems with 4000 time outputs.
We then apply the transformation to proper modes (following Sect. \ref{Laplace-Lagrange}). 
We can compute the standard deviation on this artificial "time" dimension and the median of the resulting 250 systems distribution, mimicking the \texttt{Peas} and \texttt{Far} sets considered for N-body integrations.
The resulting stds are equal to $\sim0.17$ for $\AMDe$ and $\AMDi$ and $\sim 0.11$ for the modes.

\section{Fourier expansion of secular Hamiltonian}\label{appendix:Hsec}
Writing the averaged planetary Hamiltonian as $\langle\mathcal{H}\rangle = \mathcal{H}_0 + \mathcal{H}_{\mathrm{sec}}$, with $\mathcal{H}_0$ given in Eq. \eqref{eq:H_0}, one can express $\mathcal{H}_{\mathrm{sec}}$ as
\begin{equation}
    \mathcal{H}_{\mathrm{sec}} = \sum_{\mathbf{l},\bar{\mathbf{l}},\mathbf{l}',\bar{\mathbf{l}}'} h^{\mathbf{l},\bar{\mathbf{l}},\mathbf{l}',\bar{\mathbf{l}}'}(\bm{\Lambda})\prod_{j=1}^N {x_j}^{{l_j}} \bar{x}_j^{\bar{l}_j}y_j^{l'_j}\bar{y}_j^{\bar{l}'_j}, \quad \mathbf{l},\bar{\mathbf{l}},\mathbf{l}',\bar{\mathbf{l}}' \in \mathbb{N}^N,
\end{equation}

where the coefficients $h^{\mathbf{l},\bar{\mathbf{l}},\mathbf{l}',\bar{\mathbf{l}}'}$ are real and can be computed by computer algebra \citep{Laskar_1995}.
We remind that, since the system must be invariant by reflection along the horizontal plane, $\sum_{i=0}^N l'_i + \bar{l}'_i$ is even.
The latter equation yields, using the expressions of $x_j$ and $y_j$ as functions of the action and angle canonical coordinates%
\begin{equation}
    \mathcal{H}_{\mathrm{sec}} = \sum_{\mathbf{l},\bar{\mathbf{l}},\mathbf{l}',\bar{\mathbf{l}}'} h^{\mathbf{l},\bar{\mathbf{l}},\mathbf{l}',\bar{\mathbf{l}}'}(\bm{\Lambda})\prod_{j=1}^N  I_{j}{}^{\frac{l_j+\bar{l}_j}{2}} I'_{j}{}^{\frac{l'_j+\bar{l}'_j}{2}}\mathrm{E}^{\iota\left((l_j-\bar{l}_j)\theta_j+(l'_j-\bar{l}'_j)\theta'\right)}.
\end{equation}
The degree of the expansion is defined as $\sum_j l_j+\bar{l}_j+l'_j+\bar{l}'_j$.

Let us now write $l_j+\bar{l}_j=|l_j - \bar{l}_j|+2\min(l_j,\bar{l}_j)$ and perform an (invertible) change of variable by defining $k_j$ as $k_j=l_j-\bar{l}_j$ and $\tilde{k}_j=\min(l_j,\bar{l}_j)$, equivalently we have $k'_j=l'_j-\bar{l}'_j$ and $\tilde{k}'_j=\min(l'_j,\bar{l}'_j)$.
We now have $(\mathbf{k}, \mathbf{k}')\in \mathbb{Z}^{2N}$ and
\begin{equation}
    \mathcal{H}_{\mathrm{sec}}=\sum_{\mathbf{l},\bar{\mathbf{l}},\mathbf{l}',\bar{\mathbf{l}}'} h^{\mathbf{l},\bar{\mathbf{l}},\mathbf{l}',\bar{\mathbf{l}}'}(\bm{\Lambda})\prod_{j=1}^N I_{j}{}^{\frac{|k_j|}{2}} I'_{j}{}^{\frac{|k'_j|}{2}} I_j{}^{\tilde{k}_j} I'_{j}{}^{\tilde{k}'_j} \mathrm{E}^{\iota(k_j\theta_j+k'_j\theta_j)}.
 \end{equation}
Using the proper modes canonical variables, i.e. Eq. \eqref{eq:Action_Angle}, and omitting the "$\ast$" symbols over the $\mathbf{l},\bar{\mathbf{l}},\mathbf{l}',\bar{\mathbf{l}}', k_j, k'_j, \tilde{k}, \tilde{k}'$ the secular Hamiltonian reads
\begin{equation}
    \mathcal{H}_{\mathrm{sec}}=\sum_{\mathbf{l},\bar{\mathbf{l}},\mathbf{l}',\bar{\mathbf{l}}'} h^{\ast}{}^{\mathbf{l},\bar{\mathbf{l}},\mathbf{l}',\bar{\mathbf{l}}'}(\bm{\Lambda})\prod_{j=1}^N I^\ast_{j}{}^{\frac{|k_j|}{2}} I'^{\ast}_{j}{}^{\frac{|k'_j|}{2}} I^\ast_j{}^{\tilde{k}_j} I'^{\ast}_{j}{}^{\tilde{k}'_j} \mathrm{E}^{\iota(k_j\theta^\ast_j+k'^\ast_j\theta^\ast_j)}.
 \end{equation}

At degree 2, the Hamiltonian is equivalent to the Laplace-Lagrange dynamics, integrable in the $(\mathbf{I}^\ast, \bm{\theta}^\ast)$ and $(\mathbf{I}'^\ast, \bm{\theta}'^\ast)$  pairs and reads $\mathcal{H}_{\mathrm{LL}}=-(\mathbf{g}, \mathbf{s})\cdot( \mathbf{I}^\ast, \mathbf{I}'^{\ast})$.
Let us regroup the complex conjugates, denoted $c.c.$
In doing so, the sum over  $\mathbf{k, k'}$ is done on a set $\mathcal{S} \subset \mathbb{Z}^{2N}$ such that the indices $\mathbf{k}$ (resp. $\mathbf{k'}$) and $\mathbf{-k}$ (resp. $-\mathbf{k'}$) are only counted once.
As mentioned in Sect. \ref{sec:Secular_dynamics}, one of the $s$ secular frequencies must be 0. Let us consider $s_1=0$.
Note that one can show, assuming the total angular momentum is aligned with the $z$ axis, that $y_1^\ast (t)=0, \forall t \in \mathbb{R}$.
Therefore, when expressed in the proper mode variables, non-zero terms in the Fourier decomposition must thus have $l'_1=\bar{l}'_1=0$, i.e. $k'_1 =0$. We now have 
\begin{equation}
\begin{aligned}
    \mathcal{H}_{\mathrm{sec}}= {} &\sum_{\mathbf{k,k'} \in \mathcal{S}}\mathrm{E}^{\iota(\mathbf{k}\cdot\bm{\theta}^{\ast}+\mathbf{k}'\cdot\bm{\theta}'^{\ast})}\prod_{j=1}^N I_j^{\ast\frac{|k_j|}{2}} I'^{\ast}_{j}{}^{\frac{|k'_j|}{2}}\\
    &\sum_{\tilde{\mathbf{k}}, \tilde{\mathbf{k}}'}\prod_{j=1}^N(h^\ast{}^{\mathbf{k}, \mathbf{k}'}(\bm{\Lambda})I_j^{\ast\tilde{k}_j} I'^{\ast}_{j}{}^{\tilde{k}'_j}) + c.c.,
\end{aligned}
\end{equation}
which we can write as 
\begin{equation}
\begin{aligned}
    \mathcal{H}_{\mathrm{sec}}= {} & 
    \sum_{\mathbf{k,k'} \in \mathcal{S}}\mathrm{E}^{\iota(\mathbf{k}\cdot\bm{\theta}^{\ast}+\mathbf{k}'\cdot\bm{\theta}'^{\ast})}\prod_{j=1}^N I_j^{\ast\frac{|k_j|}{2}} I'^{\ast}_{j}{}^{\frac{|k'_j|}{2}}(\tilde{h}^\ast{}^{\mathbf{k}, \mathbf{k}'}(\bm{\Lambda}, \mathbf{I}^\ast, \mathbf{I}'^{\ast})) \\
    & +c.c.
\end{aligned}
\end{equation}
For each Fourier coefficient the leading order term is obtained for $\tilde{\mathbf{k}}, \tilde{\mathbf{k}}'=(\mathbf{0,0})$ and thus reads
$\mathrm{E}^{\iota(\mathbf{k}\cdot\bm{\theta}^{\ast}+\mathbf{k}'\cdot\bm{\theta}'^{\ast})}\tilde{h}^{\mathbf{0,0}}\prod_{j=1}^N I_j^{\ast\frac{|k_j|}{2}} I'^{\ast}_{j}{}^{\frac{|k'_j|}{2}}$.
The contribution of order $2n$, $\mathcal{H}_{2n}$, to the secular hamiltonian $\mathcal{H}_{sec}$, is then obtained by selecting  $\mathbf{k}$ and $\mathbf{k'}$ such that $|\mathbf{k}| + |\mathbf{k'}|=2n$.
Finally, using this Hamiltonian function, let us define a dynamical quantity, $C$, as a linear combination of the actions, i.e.
\begin{equation}
    C_{\bm{\gamma}, \bm{\gamma'}}=(\bm{\gamma}, \bm{\gamma}')\cdot (\mathbf{I}^\ast, \mathbf{I}'^{\ast}),\quad \bm{\gamma}, \bm{\gamma}' \in \mathbb{R}^N.
\end{equation}
It is then fairly easy to compute the time derivative of $C$ using Hamilton's equations, which yields
\begin{equation}
\dot{C}_{\bm{\gamma}, \bm{\gamma'}} 
    = -\bm{\gamma}\cdot \frac{\partial \mathcal{H}_{\mathrm{sec}}}{\partial \bm{\theta}^\ast}
    -\bm{\gamma'}\cdot \frac{\partial \mathcal{H}_{\mathrm{sec}}}{\partial \bm{\theta}'^\ast}.
\end{equation}
Considering the effect of a set of $m$ distinct harmonics, the variation of $C$ reads
\begin{equation}
\begin{aligned}
\dot{C}_{\bm{\gamma}, \bm{\gamma'}, m} 
    &=  2\sum_{i=1}^m (\bm{\gamma}, \bm{\gamma}')
    \cdot (\mathbf{k}^i, \mathbf{k}'^{i})\prod_{j=1}^N I_j^{\ast\frac{|k_j|}{2}} I'^{\ast}_{j}{}^{\frac{|k'_j|}{2}}\tilde{h}^\ast{}^{\mathbf{k}, \mathbf{k}'} \\
    & \quad 
    \sin(\mathbf{k}^i\cdot\bm{\theta}^\ast+\mathbf{k}'^i\cdot\bm{\theta'}^\ast).
\end{aligned}
\end{equation}

Note that since the transformation to proper modes is equivalent to a rotation of the phase space, the \AMDe and \AMDi can be computed as the sum of $ x^\ast_j\bar{x}^\ast_j=I^\ast_j$ and of $y^\ast_j\bar{y}^\ast_j=I'^\ast_j$ respectively.
We can then define trivially \AMDe and \AMDi as $C_{\mathbf{1}, \mathbf{0}}$ and $C_{\mathbf{0}, \mathbf{1}}$ respectively and the \AMD as $C_{\mathbf{1}, \mathbf{1}}$.

\end{appendix}

\end{document}